\documentclass[]{emulateapj}

\usepackage{latexsym} 
\usepackage{amssymb}
\usepackage{amsmath}
\usepackage{graphicx}
\usepackage{tabularx}
\usepackage{booktabs}
\usepackage{subcaption}
\usepackage{hyperref}
\usepackage{fontawesome}
\usepackage{multirow}
\usepackage{xcolor}
\graphicspath{{./}{./figures/}}
\newcommand{\sameastraining}[1]{{\textcolor{gray}{#1}}}

\begin{document}

\title{From Images to Dark Matter: End-To-End Inference of Substructure From Hundreds of Strong Gravitational Lenses}
\shortauthors{Wagner-Carena et al.}
\shorttitle{From Images to Dark Matter}

\author{{Sebastian~Wagner-Carena},\altaffilmark{1,2,*}
{Jelle~Aalbers},\altaffilmark{1,2}
{Simon~Birrer},\altaffilmark{1,2}
{Ethan~O.~Nadler},\altaffilmark{3,4}
{Elise~Darragh-Ford},\altaffilmark{1,2}
{Philip~J.~Marshall},\altaffilmark{1,2}
{Risa~H.~Wechsler}\altaffilmark{1,2}}
\altaffiltext{1}{Kavli Institute for Particle Astrophysics and Cosmology, Department of Physics, Stanford University, Stanford, CA, 94305}
\altaffiltext{2}{SLAC National Accelerator Laboratory, Menlo Park, CA, 94025}
\altaffiltext{3}{Carnegie Observatories, 813 Santa Barbara Street, Pasadena, CA 91101, USA}
\altaffiltext{4}{Department of Physics $\&$ Astronomy, University of Southern California, Los Angeles, CA, 90007, USA}
\altaffiltext{*}{Corresponding Author: \email{swagnerc@stanford.edu}}

\begin{abstract}

Constraining the distribution of small-scale structure in our universe allows us to probe alternatives to the cold dark matter paradigm. Strong gravitational lensing offers a unique window into small dark matter halos ($<10^{10} M_\odot$) because these halos impart a gravitational lensing signal even if they do not host luminous galaxies. We create large datasets of strong lensing images with realistic low-mass halos, Hubble Space Telescope (HST) observational effects, and galaxy light from HST's COSMOS field. Using a simulation-based inference pipeline, we train a neural posterior estimator of the subhalo mass function (SHMF) and place constraints on populations of lenses generated using a separate set of galaxy sources. We find that by combining our network with a hierarchical inference framework, we can both reliably infer the SHMF across a variety of configurations and scale efficiently to populations with hundreds of lenses. By conducting precise inference on large and complex simulated datasets, our method lays a foundation for extracting dark matter constraints from the next generation of wide-field optical imaging surveys.
\end{abstract}


\section{Introduction}\label{sec:intro}\addtocounter{footnote}{-1}

The concordance model in cosmology, $\Lambda\text{CDM}$, includes the presence of cold, collisionless dark matter (CDM). One of the predictions of the CDM model is the presence of approximately self-similar dark matter halos ranging in mass from fractions of a solar mass to $10^{15} M_\odot$ \citep{navarro1996structure,navarro1997universal,green2004power,wang2020universal}. The formation of these structures is hierarchical, with massive halos forming from mergers and accretion of less massive halos \citep{white1978core,moore1999dark}. The CDM model further predicts that the number density of halos is inversely related to their mass, producing an abundance of low-mass ($<10^{10} M_\odot$) halos, both as gravitationally isolated structures and as `subhalos' of larger `host' halos. Many popular alternatives to CDM impact the predicted distribution, abundances, and profiles of low-mass halos \citep{bode2001halo,kaplinghat2005dark,bullock2017small,buckley2018gravitational,tulin2018dark}. Therefore, constraining dark matter halos at these small scales provides an important test of the CDM paradigm.

Over the past decade, a number of observational probes have placed constraints on low-mass dark matter halos. In the Local Group, measurements of Milky Way satellites have been used to rule out or limit the parameter space of alternative dark matter models \citep{maccio2010cold,kennedy2013constraining,nadler2019constraints,newton2021constraints,nadler2021constraints,dekker2021warm}, and early measurements of perturbations to the Milky Way's stellar streams are providing complementary constraints \citep{bonaca2019spur,banik2021evidence,banik2021novel}. At higher redshifts, Lyman-$\alpha$ forest measurements have been able to constrain dark matter models that impact small-scale structure formation in the early universe \citep{viel2013warm,irvsivc2017first,irvsivc2017new,rogers2021strong}, as have measurements of the UV galaxy luminosity function \citep{menci2016constraining,rudakovskyi2021constraints}. With the exception of stellar stream perturbations, all of these probes depend on the emission or absorption of light by baryons in the halos; connecting luminous tracers to their underlying dark matter requires accurately modeling the baryonic physics and therefore introduces large uncertainties. For Milky Way satellites, the most sensitive probe to date, modeling the galaxy--halo connection at low halo masses ($<10^9 M_\odot$), is one of the dominant observational uncertainties \citep{nadler2021constraints}. In order to further constrain small-scale structure, we require tracers that are comparable in sensitivity to existing probes but less dependent on accurate modeling of the baryonic physics.

Strong gravitational lensing is a promising low-mass halo probe, as it directly measures the gravitational signal of dark matter structure. In the case of galaxy--galaxy lenses, light from a distant source galaxy passes by a massive `main deflector' galaxy and is refocused to produce multiple images. Smaller halos along the line of sight and subhalos within the main deflector also deflect the light and can cause detectable perturbations in the lensing image. These low-mass halos\footnote{Throughout, we refer to line-of-sight halos and subhalos collectively as `low-mass halos,' and we use `subhalos' when specifically considering low-mass halos in the main lens.} can generate a signal even if they have no luminous counterparts (see \citealt{nadler2020milky} for recent upper bounds on the mass of these `dark halos' in a CDM context). The number of strong lenses available for analysis is also poised to grow rapidly, with over a thousand known lenses to date  \citep{sonnenfeld2013sl2s} and tens of thousands expected to be discovered with next-generation wide-field optical imaging surveys \citep{collett2015population}. Given modeling tools capable of extracting the low-mass halo signal, this sizable dataset provides an opportunity for state-of-the-art dark matter sensitivity.

Strong lensing studies often focus on the population of subhalos within the main lens, and specifically the abundance of subhalos per unit mass, referred to as the subhalo mass function (SHMF). Constraining the SHMF at masses below $<10^{10} M_\odot$ allows us to measure deviations from the CDM predictions for low-mass halos. Broadly speaking, there are two frameworks through which we can model the SHMF in strong lensing. The first is to attempt to detect the signal of individual subhalos in a strong gravitational lens, often called direct detection. Traditionally, this method first models a smooth main deflector and the source light. Subhalos are then added to the model, and if the improvement in the fit is sufficient---often measured through the Bayesian information criterion---the subhalos are considered detected \citep{mao1998evidence,moustakas2003detecting,koopmans2005gravitational,vegetti2009bayesian,hezaveh2013dark}. Direct detection modeling has been used to identify subhalos in three systems and place constraints on the fraction of dark matter in subhalos \citep{vegetti2010detection,vegetti2012gravitational,hezaveh2016detection,vegetti2018constraining,ccaugan2021substructure}. 
Direct detection returns a concrete picture of the exact quantity, position, and mass of the subhalos detected. However, scaling this framework is challenging; modeling only the most massive subhalos is computationally demanding and yet cannot capture the thousands of halos in the $10^7 M_\odot - 10^{9} M_\odot$ mass range. While these lower-mass subhalos cannot be individually detected, they can collectively produce an observable signal. Additionally, with only one subhalo detection per lens, constraining the SHMF requires substantial assumptions about the subhalo and line-of-sight halo populations.

The second approach, often called statistical detection, directly models the properties of the subhalo population. In statistical detection, the main deflector and source may still be explicitly modeled, but no attempt is made to detect individual subhalos. Instead, this framework measures the signal created by `all' of the subhalos and connects it to the population statistics (i.e.\ the SHMF) \citep{dalal2002direct,hezaveh2016measuring,cyr2016dark,birrer2017line,rivero2018gravitational,rivero2018power,brennan2019quantifying}. This approach reduces the number of free parameters required to describe the full mass range from thousands to dozens and can be used to directly infer the parameters that describe alternate dark matter models. However, the same SHMF can admit both subhalo configurations that are excellent and poor descriptors of the data. Additionally, because the full mass range includes thousands of subhalos, rigorously sampling the allowed configuration space of even a single SHMF is a sizable computing task. The net effect is that directly evaluating the likelihood is intractable, making traditional Markov Chain Monte Carlo inference unworkable.\footnote{Probabilistic cataloging methods conduct transdimensional Bayesian inference to overcome this limitation but remain computationally limited to the most massive $\mathcal{O}(10)$ low-mass halos \citep{brewer2016trans,daylan2018probing}}

Instead, statistical detection work has focused on simulation-based inference, a family of methods that circumvent the need for a tractable likelihood by leveraging access to a simulator. The most well known of these methods is approximate Bayesian computation (ABC) \citep{rubin1984bayesianly,beaumont2002approximate}. In ABC, simulated data is generated by sampling from a prior on the parameter space. If the summary statistics of the simulated data approximately match the observed data, the sample is kept. With a sufficiently strict matching criteria, ABC returns a faithful sampling of the likelihood. In the strong lensing context, ABC has been used to constrain warm dark matter models, the halo mass--concentration relation, and the primordial power spectrum \citep{birrer2017lensing,gilman2020warm,gilman2020constraints,gilman2021primordial,nadler2021dark}. However, ABC has its own drawbacks as an inference methodology. First, for image datasets, ABC requires reducing the images to low-dimensional summary statistics that discard information and therefore limits the constraining power of the data. Second, ABC inference does not scale to large datasets; each observed lens requires an independent inference chain with hundreds of thousands of personalized simulations. 

In light of the drawbacks to ABC, modern simulation-based inference has shifted toward using neural networks as density estimators. In this approach, a network is trained on a single training set to predict either the posterior \citep{lueckmann2017flexible}, the likelihood \citep{papamakarios2019sequential}, or a likelihood ratio \citep{mohamed2016learning} given an input datapoint. Unlike ABC, the density estimator can take advantage of the full information content of the data. The density estimator is also scalable: after the initial training, the cost of conducting inference on a new lens is negligible. In strong lensing, \cite{brehmer2019mining} have demonstrated that a likelihood ratio estimator can accurately extract the signal of subhalos on a population of 100 simulated lenses, and recent work has also applied simulation-based inference techniques to the direct detection framework \citep{PhysRevD.101.023515,ostdiek2020detecting,ostdiek2020extracting,lin2020hunting,coogan2020targeted}. Outside of the subhalo context, several studies have shown the ability of neural density estimators to constrain the parameters of the main deflector \citep{perreault2017uncertainties,wagner2021hierarchical,pearson2021strong} and to infer the Hubble constant from strong lensing time-delay measurements \citep{park2021large}.

The recent literature suggests that applying neural density estimators to strong lensing images will allow us to place tight constraints on alternatives to CDM. Despite this, neural density estimators of subhalo parameters have never been applied to observed strong lensing images. The principal limitation is the simulations. The existing proof-of-concept studies make a number of simplifications: these include smooth source models, ignoring line-of-sight halos, and simplistic detector responses. Using networks trained on simplified simulations is likely to produce biased inference, and it may be that adding these complexities will wash out the constraining power of the networks. With an eye toward pushing neural density estimator techniques toward the data, we set out to answer the following questions: given simulations of strong lensing images generated using realistic assumptions for the sources, subhalos, line-of-sight halos, and Hubble Space Telescope (HST) observational effects, can simulation-based inference with neural networks constrain the subhalo population on individual lenses? Specifically, can the network constrain the normalization of the SHMF? And then, can a network trained to make predictions on individual lenses be used to hierarchically infer the SHMF of a population of lenses?

\begin{figure*}
    \centering
    \includegraphics[scale=0.35]{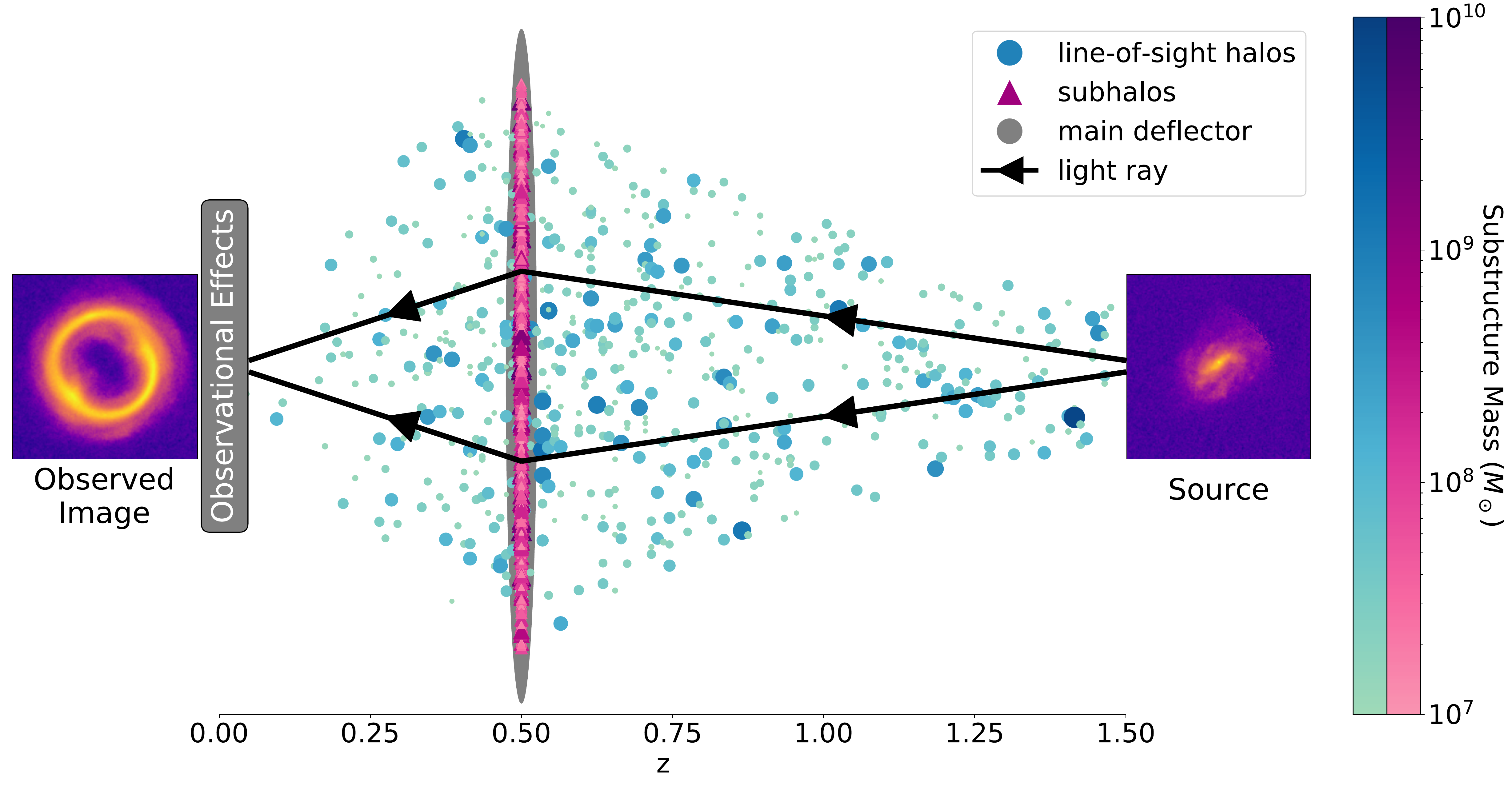}
    \caption{Schematic summary of the ingredients of our strong gravitational lensing simulation. The light rays leave the source (Section \ref{sec:source}), are perturbed by the line-of-sight halos (Section \ref{sec:los}), are bent by the main deflector (Section \ref{sec:main_def}) and further perturbed by the subhalos (Section \ref{sec:subhalos}) and remaining line-of-sight halos to finally be measured by our detector. On top of the raytracing, there are also observational effects from the detector (Section \ref{sec:obs_and_num}) that further distort the image. The final observed image is shown on the far left.}
    \label{fig:sim_sum}
\end{figure*}

In this paper, we address these questions by using realistic, complex simulations of galaxy--galaxy strong lenses to train and test a neural posterior estimator of the SHMF normalization. We pull our sources from observed galaxies, marginalize over well-motivated theoretical uncertainties when sampling our low-mass halos, and include the effects of the HST imaging pipeline. We test our neural posterior estimator's ability to accurately extract the SHMF signal on individual simulated lenses, including when the underlying sources are not seen during training. For this study, we confine our analysis to the $\Lambda\text{CDM}$ model with cosmological parameters fixed to the best-fit values from \cite{aghanim2020planck}. We combine our network with a hierarchical inference pipeline to reconstruct the SHMF normalization for test sets that are statistically distinct from our training set. We find that our simulation-based inference approach is capable of accurately and efficiently inferring the statistic of a lensing population with hundreds of observations. Our method is sensitive to SHMF normalizations spanning over an order of magnitude and can account for systematic offsets between the distribution of training and test populations.

This paper is organized as follows. In Section \ref{sec:simulation} we discuss the simulated strong lensing images used in this work, with a particular focus on the choices surrounding the low-mass halos, the sources, and the HST observational effects. We then introduce our simulated datasets, neural posterior estimator, and hierarchical inference scheme in Section \ref{sec:methods}. We demonstrate the ability of the trained model to predict the normalization of the SHMF on individual lenses in Section \ref{sec:perf_train}. In Section \ref{sec:recon_shmf} we run our model on twenty sets of lensing images, each with a self-similar SHMF, and explore the constraints we can extract with HST-quality photometry. We discuss the limitations of our analysis and potential future extensions in Section \ref{sec:discuss}. Finally, in Section \ref{sec:conc} we summarize our results and discuss the implications of our work for the future of dark matter constraints with strong lensing.

As part of this publication, we are releasing our strong lensing simulation package \textsc{paltas} \footnote{\url{https://github.com/swagnercarena/paltas}}. The package builds on the simulation code \textsc{lenstronomy}\footnote{\url{https://github.com/sibirrer/lenstronomy}} \citep{birrer2018lenstronomy,birrer2021lenstronomy} to produce large and realistic strong lensing image datasets. The code has been designed to be modular, scalable, and easy to configure. Our hope is that it will provide an effective tool for the future development of simulation-based inference in strong lensing. The \textsc{paltas} repository includes all of the code and dependencies necessary to reproduce the results in this paper along with a set of comprehensive \textsc{Jupyter} notebooks that help familiarize users with the code.

\section{Simulation Methods}\label{sec:simulation}

Extracting unbiased information with simulation-based inference requires representative simulations of the data. In the following subsections, we describe the parameterizations we use to simulate realistic strong lensing images that incorporate our existing theoretical uncertainties. We break our simulation choices down into five components: the main deflector (Section \ref{sec:main_def}), the subhalos of the main deflector (Section \ref{sec:subhalos}), the line-of-sight halos (Section \ref{sec:los}), the source (Section \ref{sec:source}), and the observational and numerical parameters for the simulation (Section \ref{sec:obs_and_num}). A schematic representation of the model can be found in Figure \ref{fig:sim_sum}. While the distribution of parameter values we use varies between the training, validation, and test sets, the parameterizations themselves remain consistent throughout this work. Table \ref{tab:dist} lists all parameters of our model, along with their distribution on each image set. 

For many of the cosmology calculations in our simulations we use the \textsc{colossus}\footnote{\url{https://bitbucket.org/bdiemer/colossus}} package \citep{diemer2018colossus}. For the lensing calculations we use the lens modeling package \textsc{lenstronomy}.

\subsection{Main Deflector}\label{sec:main_def}

In our simulations, the main deflector is a power-law elliptical mass distribution (PEMD) profile with external shear. The PEMD \citep{kormann1994isothermal,barkana1998fast} profile is described by the convergence:

\begin{align}\label{eq:pemd}
    \kappa (x,y) &= \frac{3-\gamma_\text{lens}}{2}\left( \frac{\theta_E}{\sqrt{q_{\text{lens}}x^2 + y^2/q_{\text{lens}}}}\right)^{\gamma_\text{lens}-1}.
\end{align}

Here $\gamma_\text{lens}$ is the logarithmic slope, $\theta_E$ the Einstein radius, and $q_{\text{lens}}$ the axis ratio of the lens. The profile described by Equation~\ref{eq:pemd} assumes that the coordinate system for $x$ and $y$ is defined along the major and minor axes of the deflector. Therefore, there are three remaining parameters to fully describe the profile: the main deflector center position $(x_{\text{lens}},y_{\text{lens}})$ and the main deflector rotation angle
$\phi_\text{lens}$. The additional external shear component is described by an orientation angle $\phi_\text{ext}$ and modulus $\gamma_\text{ext}$ \citep{keeton1997shear}.

The angles $\phi_\text{lens}$ and $\phi_\text{ext}$ are cyclic parameters, which would complicate inference. We therefore work in the eccentricity / Cartesian coordinates for our ellipticity / shear:

\begin{align}
    e_1 &= \frac{1-q_{\text{lens}}}{1+q_{\text{lens}}} \cos (2 \phi_\text{lens})\\
    e_2 &= \frac{1-q_{\text{lens}}}{1+q_{\text{lens}}} \sin (2 \phi_\text{lens})\\
    \gamma_1 &= \gamma_\text{ext} \cos (2 \phi_\text{ext}) \\
    \gamma_2 &= \gamma_\text{ext} \sin (2 \phi_\text{ext}).
\end{align}

\footnotetext{Our definition of $M_{200c}$ uses the critical density at the redshift of the subhalo, not the critical density at redshift zero.}

Finally, the main deflector has redshift $z_\text{lens}$ and host mass $m_\text{host}$. The host mass uses the $M_{200c}$ definition \citep{white2001mass}\footnotemark[\value{footnote}]. There does not exist an exact mapping from the host mass to the Einstein radius, therefore the two values are left uncorrelated in our simulations.

\subsection{Subhalos}\label{sec:subhalos}

The subhalos of the main deflector in our simulations follow the parameterization introduced by \cite{gilman2020warm} with some slight modifications. We draw our subhalos from the following mass function:

\begin{align}\label{eq:shmf}
    \frac{d^2 N_\text{sub}}{dA \ dm_\text{sub}} &= \Sigma_\text{sub} \frac{m_\text{sub}^{\gamma_\text{sub}}}{m_{\text{pivot,sub}}^{\gamma_\text{sub}+1}},
\end{align}
where $\Sigma_\text{sub}$ is the normalization of the SHMF, $m_\text{sub}$ is the subhalo mass using the $M_{200c}$ definition \citep{white2001mass}\footnotemark[\value{footnote}], $dA$ is the differential area element, and $m_{\text{pivot,sub}}$ is the pivot mass. We render subhalos within the mass range $[m_{\text{min,sub}},m_{\text{max,sub}}]$\footnotetext{We assume that more massive halos would host sufficient baryons to be visible in our images and therefore modeled individually. The lower limit is set below the sensitivity of our inference (see Appendix \ref{app:params}). }\footnotemark[\value{footnote}]\newcounter{footsave}\setcounter{footsave}{\value{footnote}}.

The SHMF as written contains no explicit dependence on host properties. Any scaling by, for example, the host mass or redshift has been absorbed into our definition of $\Sigma_\text{sub}$. We expect our network to be sensitive to the projected number of subhalos in the main deflector, which for a fixed slope $\gamma_\text{sub}$ is best captured by $\Sigma_\text{sub}$. For a realistic/observed population of lenses, we therefore expect our framework to return a distribution of $\Sigma_\text{sub}$ values that must be interpreted in the context of a model with host-dependent scaling.

The subhalos themselves are modeled as a truncated NFW radial density profile \citep{baltz2009analytic}. The profile can be defined in terms of a mass, $m_\text{sub}$, a concentration, $c_\text{sub}$, and a truncation radius $r_t$. A detailed discussion of how these are drawn in our simulation can be found in Appendix \ref{app:subhalo}. For the positions of the subhalos, we follow \cite{gilman2020warm}. Specifically, outside of the host's scale radius, $r_{s,\text{host}}$, the subhalos follow the host's mass profile; within $r_{s,\text{host}}$, the subhalos are uniformly distributed. To keep the simulations numerically tractable without altering the signal, we render subhalos within a projected radius of $3 \theta_E$, where $\theta_E$ is the Einstein radius of the main deflector. The $z$-coordinates of the subhalos are also constrained to be within the interval $[-R_{200c},R_{200c}]$. Here, $R_{200c}$ is the smallest radius such that the host halo's enclosed mass has a mean density of 200 times $\rho_\text{crit}(z)$, the critical density of the universe at redshift $z$. Outside of this radius, potential halos are considered line-of-sight halos and accounted for in the two-point halo correlation (see Section \ref{sec:los}).  

\subsection{Line-of-Sight Halos}\label{sec:los}

Historically, several studies of galaxy--galaxy strong lenses have ignored the contributions from line-of-sight halos \citep{vegetti2010detection,vegetti2012gravitational}. However, \cite{despali2018modelling} and \cite{csengul2020quantifying} have shown that, for certain lensing configurations, line-of-sight halos can produce a signal on par with the subhalos of the main deflector. In fact, \cite{despali2018modelling} and \cite{ccaugan2021substructure} demonstrate that one of the two existing subhalo detections can be better explained by a line-of-sight perturber. Therefore, while we do not explore the ability of strong lensing to constrain line-of-sight halos in this work, we do model the line-of-sight halos in our simulations, self-consistently within the same CDM framework. This allows us to marginalize over their uncertainties and calculate their impact on the SHMF signal, as has been done by previous work \citep{gilman2020warm,gilman2020constraints}.\footnote{It is also possible to use our simulations to marginalize over the subhalo contribution to constrain the line-of-sight halo population, although we do not do that here.}

As with the subhalos, the line-of-sight halos in our simulations closely follow the parameterization introduced in \cite{gilman2020warm}. We draw our line-of-sight halos in discrete redshift bins ranging from a minimum redshift, $z_\text{min,los}$, to the redshift of the source, $z_\text{source}$. The width of the redshift bins is set by the parameter $\Delta_{z,los}$. Within each redshift bin, we draw our line-of-sight halos from a modified Sheth--Tormen halo mass function \citep{sheth2001ellipsoidal}:

\begin{multline}
    \frac{d^2 N_\text{los}}{dV \ dm_\text{los}} = \delta_\text{los}(1+\xi_\text{2 halo}(r,m_\text{host},z_\text{host})) \times \\ \left[\frac{d^2 N_\text{los}}{dV \ dm_\text{los}}\right]_\text{ST}.
\end{multline}
Here $\delta_\text{los}$ is a scaling parameter that accounts for uncertainties in overall normalization of the line-of-sight mass function, $\xi_\text{2 halo}(r,m_\text{host},z_\text{host})$ is a contribution from the two-point halo correlation function, and $\left[\frac{d^2 N_\text{los}}{dV \ dm_\text{los}}\right]_\text{ST}$ is the traditional Sheth--Tormen halo mass function. Each of these components is described in further detail in Appendix \ref{app:los}. We render our line-of-sight halos in the mass range $[m_\text{min,los},m_\text{max,los}]$\footnotemark[\value{footsave}].

The line-of-sight halos themselves are parameterized by an NFW profile \citep{navarro1997universal} with the radial density profile:

\begin{align}
    \rho_\text{NFW}(r) = \frac{\rho_\text{los}}{r/r_{s,\text{los}} \left( 1 + r/r_{s,\text{los}}\right)^2}.
\end{align}
Here $\rho_\text{los}$ is the amplitude of the NFW density profile in units of $M_\odot/kpc^3$, $r$ is the radial position in units of kpc, and $r_{s,\text{los}}$ is the scale radius in units of kpc. As with the subhalo profile, the scale radius, $r_{s,\text{los}}$, and amplitude of the density profile, $\rho_\text{los}$, are calculated from the mass, $m_\text{los}$, and concentration, $c_\text{los}$, of the line-of-sight halo. The mass--concentration relation being used is fully detailed in Appendix \ref{app:subhalo}.

The $z$-coordinate of each line-of-sight halo is set by the redshift slice it is in, and the $x$-coordinate and $y$-coordinate are bound within a double cone. The cone is defined by an opening angle from the observer, $\theta_\text{los}$, and peaks in radius at the main-deflector redshift $z_\text{lens}$. From there the cone closes with an angle set by the requirement that the radius of the cone at the source redshift, $z_\text{source}$, must be 0.2 of the radius at $z_\text{lens}$. Within the cone, the $x$-coordinate and $y$-coordinate of the line-of-sight halo are sampled uniformly, as we show in Figure \ref{fig:sim_sum}.

Finally, we must also add negative convergence sheets to the lensing potential to cancel the mean expected convergence from the line-of-sight halos. This procedure avoids rendering lines of sight that are systematically overdense relative to the matter density of the universe \citep{birrer2017line}. At a fixed redshift, the deflection angles generated by our line-of-sight halos add linearly to one another. Therefore, we calculate the mean expected convergence by generating the deflection angles of an NFW halo with mean mass and concentration and then convolving the deflection angles with the uniform disc onto which we render our line-of-sight halos. The negative of these convolved deflection angles then defines the convergence sheet that must be added. Because this operation is expensive, we use a slightly wider binning in redshift, $\Delta_\text{z,correction}$, for the convergence sheet calculations. We choose $\Delta_\text{z,correction}$ so that the error in the average convergence is negligible.

\subsection{Source}\label{sec:source}

\begin{figure}
    \centering
    \includegraphics[scale=0.20]{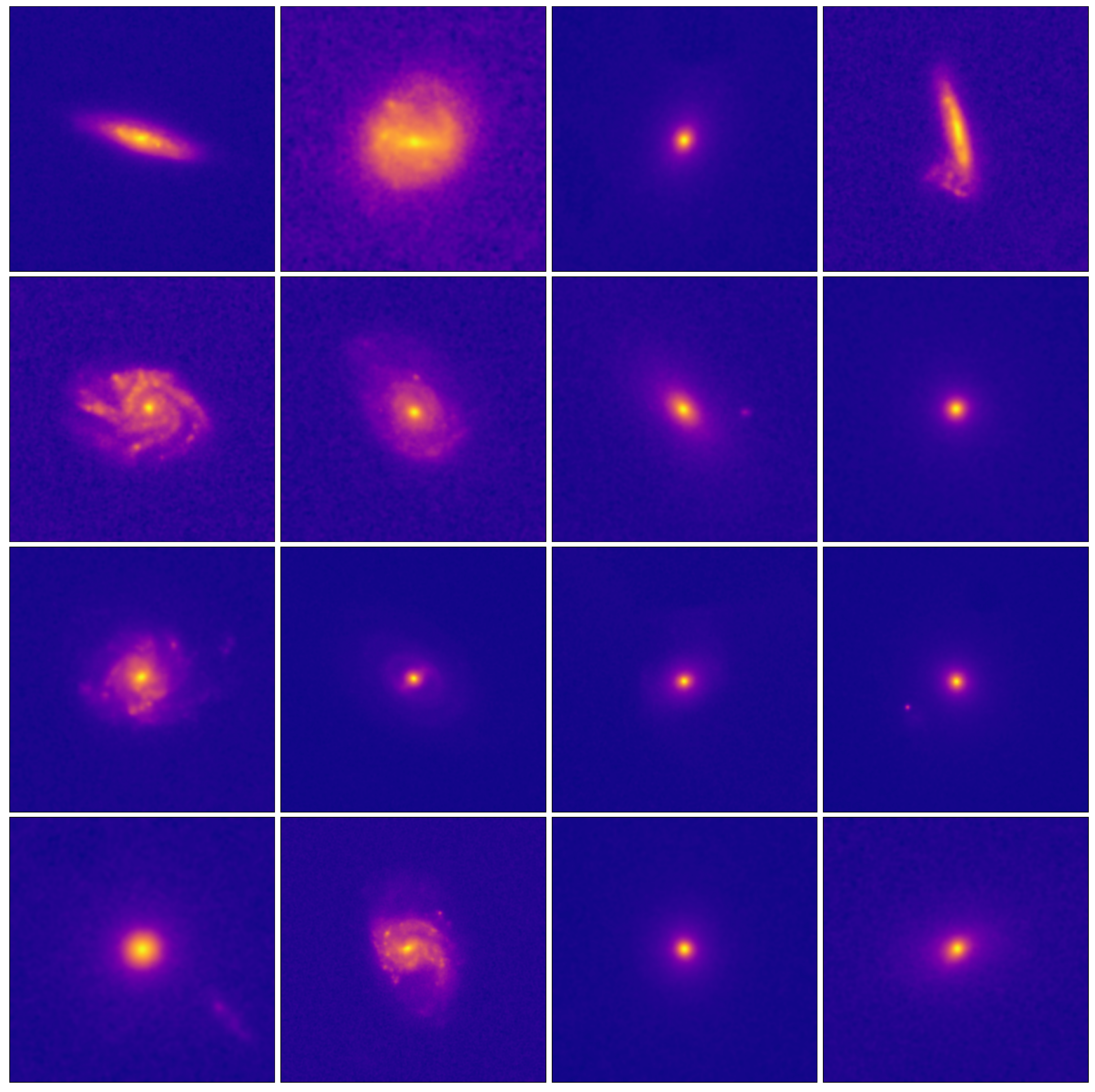}
    \caption{Sample of the HST's COSMOS \citep{koekemoer2007cosmos} galaxy images being used as sources by the \textsc{paltas} pipeline. The cutouts shown here have been generated using the procedure discussed in \cite{mandelbaum2014third} as well as the additional cuts described in Section \ref{sec:source}.}
    \label{fig:COSMOS_sources}
\end{figure}

Previous studies using neural posterior estimators for strong lensing inference have focused on simple, parametric source models \citep{perreault2017uncertainties,brehmer2019mining,wagner2021hierarchical}. Here, the sources in our simulation are drawn from 2,262 real galaxy images taken by the HST COSMOS survey \citep{koekemoer2007cosmos}. The images were taken using the HST Advanced Camera for Surveys (ACS) \citep{acs_inshandbook} between October 2003 and June 2005 using the F814W filter. From this larger survey, we use the subsample of postage stamp images generated for the GREAT3 gravitational lensing challenge \citep{mandelbaum2012precision,mandelbaum2014third} and distributed with the package \textsc{GalSim}\footnote{The image database can be found at \url{https://github.com/GalSim-developers/GalSim/wiki/RealGalaxy\%20Data}} \citep{rowe2015galsim}. A more detailed discussion of this dataset can be found in \cite{mandelbaum2014third}, and we summarize the important points in Appendix \ref{app:source}. Our pipeline takes this image catalog and imposes a few additional selection cuts to ensure that we are only using well-resolved galaxies: a minimum cutout size in pixels, $\text{size}_\text{min,pix}$, a faintest apparent magnitude, $\text{mag}_\text{faint}$, a maximum redshift, $z_\text{catalog,max}$, and a minimum estimated half-light radius, $r_{1/2}$, in units of pixels. After inspecting all images that passed these cuts, we removed 110 images without a well-imaged galaxy. These mainly showed strong point sources, or had much of their light masked out by preprocessing algorithms. We kept images with multiple blended or nearby galaxies. A sample of the 2,262 remaining source images can be seen in Figure \ref{fig:COSMOS_sources}. The values we use for the cut parameters can be found in Table \ref{tab:dist}.

Our model for the light is then a linear interpolation of one of these 2,262 images. This interpolation introduces three additional degrees of freedom to our source model: the rotation angle of the source, $\phi_\text{source}$, and the $x$- and $y$-coordinates of the source, $x_\text{source},y_\text{source}$. For this source model, we keep the absolute luminosity and physical size of the galaxy fixed to what is observed in the COSMOS images. To do this, we first scale both the angular size and measured flux of the galaxy to the values that would be measured at the source redshift, $z_\text{source}$. Then we convert from the electron count units of the F814W filter on the ACS detector to the electron count units of our target detector using the offset in the AB magnitude zeropoints. For the F814W filter on ACS images we assume an AB zeropoint of 25.95, which is the average zeropoint over the COSMOS survey period \citep{koekemoer2007cosmos,mandelbaum2014third,acs_inshandbook}.

%%%%%%%%%%

\subsection{Observational and Numerical Parameters}\label{sec:obs_and_num}

Our simulated lens images are made assuming observations by the HST Wide Field Camera 3 (WFC3) UVIS channel with the F814W filter. The low-mass halo signal we want to detect is nearly at the level of the noise. Therefore, the distortions and correlated noise generated by the HST camera and pipeline are important systematics. Our simulations take care to include observational effects that are as realistic as possible. 

We use the UVIS pixel-size of 0.040 arcsec/pixel\footnote{Note that due to drizzling, the final resolution of our images is 0.030 arcsec/pixel, not 0.040 arcsec/pixel.} (\citealt{wfc3_datahandbook}, section 1.1), a CCD gain of 1.58 (\citealt{wfc3_datahandbook}, section 5.1.1), and a read noise of $3 e^-$ (\citealt{wfc3_datahandbook}, section 5.1.2). For the AB magnitude zero point we take the 2020 corrected measurement of 25.127 \citep{calamida2021new}\footnote{Summary of results available on the \href{https://www.stsci.edu/hst/instrumentation/wfc3/data-analysis/photometric-calibration/uvis-photometric-calibration}{STScI website}.}. To calculate the expected combined sky noise and dark current we use the HST Exposure Time Calculator\footnote{\url{https://etc.stsci.edu/etc/input/wfc3uvis/imaging/}}, which gives a total brightness of 21.83 $\text{magnitude} / \text{arcsec}^2$ for average zodiacal light and earth shine conditions.

Our point-spread function (PSF) is pulled from the WFC3 PSF Databse \citep{dauphin2021wfpc2}\footnote{A selection of PSFs can be found at \url{https://www.stsci.edu/hst/instrumentation/wfc3/data-analysis/psf}, and the specific PSF map used by \textsc{paltas} can be found at \url{https://github.com/swagnercarena/paltas/blob/main/datasets/hst_psf/emp_psf_f814w.fits}.}. These PSF models have been empirically constructed from dithered observations of the star cluster Omega Centauri and are supersampled by a factor of 4 relative to the pixel size of the CCD. The modeling allows for spatial variation of the PSF across the detector, and we select the best empirical model for the center of UVIS Chip 1. We degrade the PSF to a supersampling factor of 2 (i.e.\ twice the resolution of the CCD) in order to match the resolution at which we simulate our light rays.

The \textsc{paltas} simulations we use in this work also account for the effects of HST's drizzling pipeline. HST science images are not just the raw detector output. Rather, the detector measurements are first run through the \textsc{DrizzlePac} pipeline\footnote{\url{https://stsci.edu/scientific-community/software/drizzlepac.html}}. Besides standard corrections for the sky background and cosmic rays, the \textsc{DrizzlePac} pipeline is also responsible for correcting the geometric distortions caused by the tilt of the focal plane relative to the optical axis of the detector. To do this, the \textsc{DrizzlePac} pipeline uses the drizzle algorithm \citep{fruchter2002drizzle} to combine the information from multiple dithered exposures into a final geometrically corrected image. Because the algorithm uses multiple exposures that are offset by subpixel intervals, the information content is at a higher resolution than the native resolution of the detector. Therefore, the images output by the drizzle algorithm can be at a smaller pixel scale than the detector. In fact, some amount of supersampling is important for minimizing both the aliasing and shape variation in the corrected image \citep{rhodes2007stability}. From the simulation standpoint, the drizzling pipeline imposes two major challenges. First, the PSF is a property of the image in the detector plane, and therefore empirical PSF measurements are based on the resolution of the detector not the resolution of the output of the drizzle algorithm. Second, the information in each pixel of the detector plane is spread out among several pixels in the final drizzled image. This leads to correlated noise that can mimic the small scale deflection signal produced by low-mass halos.

In order to best capture the effect the \textsc{DrizzlePac} pipeline will have on our sensitivity to the SHMF, \textsc{paltas} uses the \textsc{drizzle}\footnote{\url{https://github.com/spacetelescope/drizzle}} software package to run the raytraced images through the drizzling algorithm. Here, raytracing refers to the process of simulating the path of the emitted light rays from the source to the observer. The detailed steps can be found in Appendix \ref{app:drizz}. The addition of the noise and the PSF convolution happen before the drizzling, giving the final image both a realistically distorted PSF and correlated noise\footnote{\cite{ding2018time} captured the effects of the drizzle algorithm using a similar pipeline.}. For each of the four dithered images used in the \textsc{drizzle} pipeline, we simulate a 23 minute exposure, equivalent to a total of 2 orbits of HST time per lens (\citealt{wfc3_inshandbook}, section 10.4.4). We chose a postage stamp of 128x128 pixels in the detector plane so that the final size of our images after drizzling is $170\times170$ pixels or $5.1\arcsec \times 5.1\arcsec$. Additionally, a mask of radius $0.5\arcsec$ is placed at the center of the image to represent the region that must be discarded due to a lens light subtraction process.

\section{Inference Methods}\label{sec:methods}

\begin{figure*}
    \centering
    \includegraphics[scale=0.86]{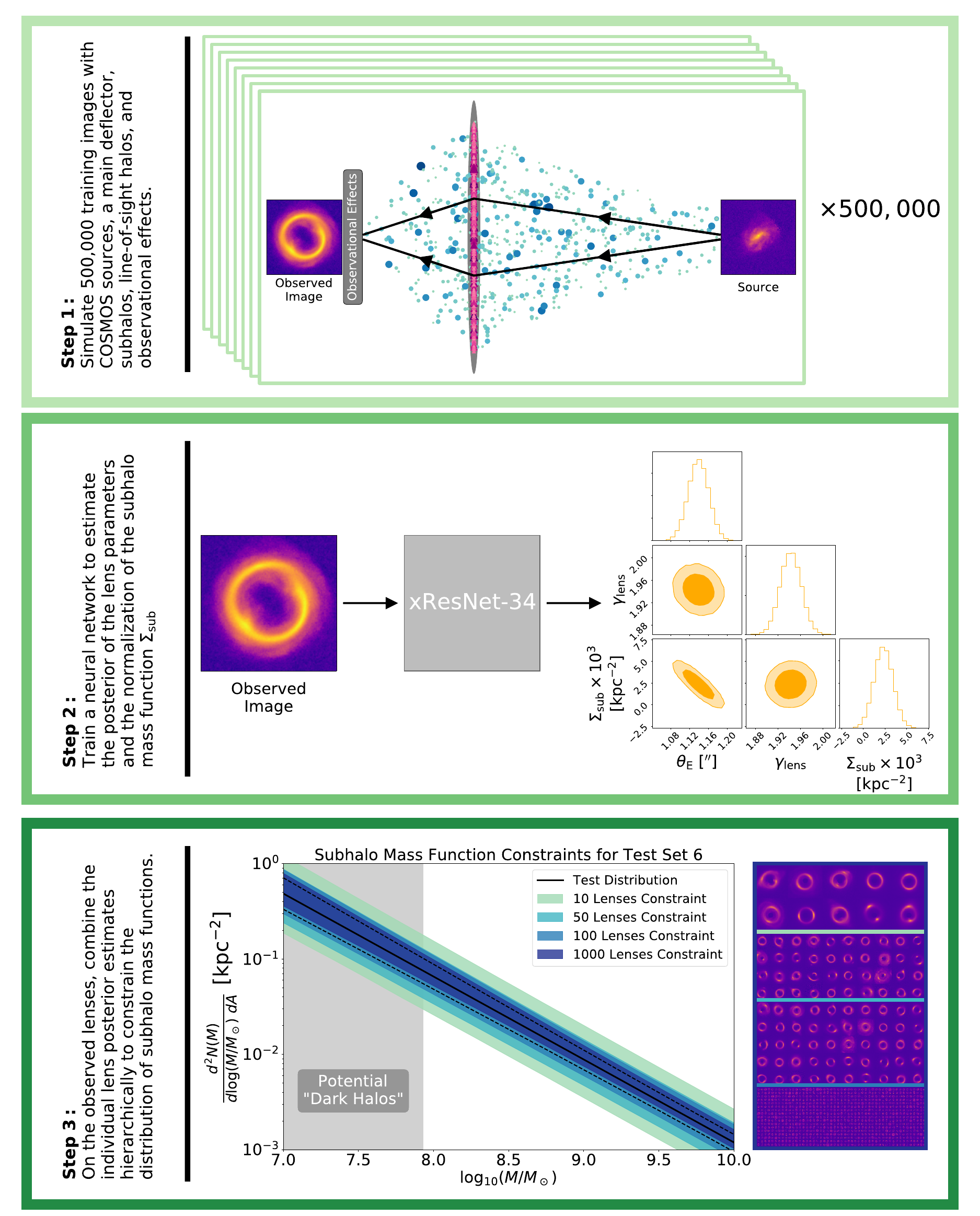}
    \caption{Summary of our simulation-based inference pipeline. The top plot depicts the first step --- generating 500,000 simulated lenses that include realistic sources, dark matter structure, and observational effects (see Section \ref{sec:sim_datasets}). Second, we use these 500,000 lenses to train a neural posterior estimator. This network constrains the lensing parameters on individual strong lensing images; the middle plot shows the posteriors returned by the trained network for a single lens image (see Section \ref{sec:nn_model}). Finally, we combine these individual lens posteriors hierarchically to constrain the subhalo mass function for a population of lenses. The bottom plot shows the subhalo mass function constraints derived from lens images drawn from a different lens parameter, source, and subhalo mass function distribution than the training set (see Section \ref{sec:hi_inf}).}
    \label{fig:pipeline}
\end{figure*}

Our simulation-based inference pipeline has three steps:
\begin{enumerate}
\item Generate a 500,000 image training set with wide distributions on the parameters of interest (Section \ref{sec:sim_datasets}). 
\item Train a neural density estimator to estimate the posterior distribution of the parameters of individual lenses (Section \ref{sec:nn_model}). 
\item Hierarchically combine the single-image posteriors of a large population of lenses to infer the distribution of subhalo mass function normalizations (Section \ref{sec:hi_inf}). 
\end{enumerate}
The following three sections go into more detail on each of these steps, and the pipeline is further summarized in Figure \ref{fig:pipeline}.

\subsection{Simulated Datasets}\label{sec:sim_datasets}

For the results presented in this work, we have simulated twenty two datasets: one training dataset, one validation dataset, and twenty test datasets. The parameter distributions for the training, validation, and test datasets can be found in Table \ref{tab:dist}. The training dataset is used to fit the network parameters and consists of 500,000 strong gravitational lensing images following the simulation choices outlined in Section \ref{sec:simulation}. This includes a PEMD main deflector with external shear, subhalos of the main deflector, halos along the line of sight, sources pulled directly from HST COSMOS observations, and the observational properties of HST's WFC3. The validation dataset is drawn from the same distributions as the training dataset, but includes only 1000 images. The validation dataset is used to evaluate the trained network's performance on individual lenses. 

The twenty test datasets have been constructed to test our pipeline's ability to constrain a range of SHMF normalizations. Each test dataset contains 1,000 images generated using the same models as the training dataset; however, the distribution of the eight parameters that describe the main deflector and the parameter controlling the normalization of the subhalo mass function have been modified. These are the nine parameters that our network infers (see Section \ref{sec:nn_model}). The mean SHMF normalizations linearly span the interval $\Sigma_\text{sub} \in [2\times 10^{-4} \ \text{kpc}^{-2} ,4\times 10^{-3}\ \text{kpc}^{-2}]$. For the eight main deflector parameters, the test distributions are narrower than the training dataset parameter distributions and shifted uniformly by half a sigma. These shifts will be used to confirm that our inference of the SHMF normalization is not biased by the choices of the training set. We will discuss this further in Section \ref{sec:hi_inf} and Section \ref{sec:recon_shmf}. 

The COSMOS catalog parameters have been chosen to use galaxies that are relatively nearby and better resolved. We impose these cuts because we are interested in exploring the degeneracy between galaxy morphology and the subhalo signal. However, this also means that for this analysis we assume the population of large, nearby galaxies resembles the high-redshift population. In Section \ref{sec:discuss} we will discuss alternatives to this assumption for future analysis. The cuts we impose are the same for the training, validation, and test datasets, however of the 2262 galaxies that survive our cuts, 2163 are reserved for training and 99 are reserved for validation and test images. Therefore, at test time, we are conducting inference on sources the network has never seen.

For the training, validation, and test datasets, the subhalo mass function slope is drawn from a wide uniform distribution, to represent marginalizing over theoretical uncertainties estimated using cosmological CDM simulations and semi-analytic models \citep{benson2020normalization}. Similarly, we account for theoretical uncertainty in the line-of-sight structure mass function by drawing its normalization uniformly between zero (no line-of-sight structure) and twice the theoretical expectation. We also draw our mass--concentration relation parameters from a wide uniform distribution, to cover values favored by modern large-scale structure simulations \citep{bullock2001profiles,prada2012halo,ludlow2016mass,diemer2019accurate}. Finally, we draw the mass--concentration relation scatter from a uniform distribution that covers the results of cosmological CDM simulations \citep{dutton2014cold,diemer2019accurate,diemer2015universal}.

Both the SHMF normalization and the line-of-sight mass function normalization are allowed to take on negative values in the distributions we have assigned. These negative values are not physically meaningful---when simulating, we treat a negative normalization as though it were zero. We chose to include these negative values rather than truncate our normal distributions because it allows us to conduct our hierarchical inference analytically\footnote{We tried inferring the logarithm of these normalization parameters instead, but saw increased non-Gaussianity in the joint posteriors with other parameters, such as $\theta_E$.} (see Section \ref{sec:hi_inf}).

The remaining parameters presented in Table \ref{tab:dist} are discussed in Appendix \ref{app:params}.

\begin{table*}
    \centering
    \renewcommand{\arraystretch}{1.3}
    \begin{tabularx}{\textwidth}{@{} p{6.4cm} X p{5.8cm} @{}}
    \toprule
    \textbf{Component} & \textbf{Training/Validation Distribution} & \textbf{$\Sigma_\text{sub}$ Test Distributions 1,...20} \\
    \midrule
    \raggedright \textbf{Main Deflector --- Section \ref{sec:main_def}} &  \\
    x-coordinate lens center ($\arcsec$) & $x_\text{lens} \sim \mathcal{N}(\mu: 0, \sigma: 0.16)$ & $x_\text{lens} \sim \mathcal{N}(\mu: \text{Unif}(-0.08,0.08)^\star, \sigma: 0.016)$\\
    
    y-coordinate lens center ($\arcsec$) & $y_\text{lens} \sim \mathcal{N}(\mu: 0,\sigma: 0.16)$ & $y_\text{lens} \sim \mathcal{N}(\mu: \text{Unif}(-0.08,0.08)^\star, \sigma: 0.016)$\\
    
    Einstein Radius ($\arcsec$) & $\theta_E \sim \mathcal{N}(\mu: 1.1, \sigma: 0.15)^\dag$ & $\theta_E \sim \mathcal{N}(\mu: \text{Unif}(1.025,1.175)^\star, \sigma: 0.015)^\dag$\\
    
    Power-law slope & $\gamma_\text{lens} \sim \mathcal{N}(\mu: 2.0,\sigma: 0.1)^\dag$ & $\gamma_\text{lens} \sim \mathcal{N}(\mu: \text{Unif}(1.95,2.05)^\star,\sigma: 0.01)^\dag$ \\
    
    \raggedright x-direction ellipticity eccentricity & $e_1 \sim \mathcal{N}(\mu: 0,\sigma: 0.1)$ & $e_1 \sim \mathcal{N}(\mu: \text{Unif}(-0.05,0.05)^\star,\sigma: 0.01)$ \\
    
    \raggedright xy-direction ellipticity eccentricity & $e_2 \sim \mathcal{N}(\mu: 0,\sigma: 0.1)$ & $e_2 \sim \mathcal{N}(\mu: \text{Unif}(-0.05,0.05)^\star,\sigma: 0.01)$\\
    
    Main halo critical mass $(M_\odot)$ & $m_\text{host} = 10^{13}$ & \sameastraining{$m_\text{host} = 10^{13}$} \\
    
    Main halo redshift & $z_{\text{lens}} = 0.5$ & \sameastraining{$z_{\text{lens}} = 0.5$} \\
    
    x-direction shear & $\gamma_1 \sim \mathcal{N}(\mu: 0,\sigma: 0.05)$ & $\gamma_1 \sim \mathcal{N}(\mu: \text{Unif}(-0.025,0.025)^\star,\sigma: 0.005)$\\
    
    xy-direction shear & $\gamma_2 \sim \mathcal{N}(\mu: 0,\sigma: 0.05)$ & $\gamma_2 \sim \mathcal{N}(\mu: \text{Unif}(-0.025,0.025)^\star,\sigma: 0.005)$\\
    
    \midrule
    \textbf{Mass--concentration --- Appendix \ref{app:subhalo}} &  \\
    \raggedright Concentration normalization & $c_0 = \text{Unif}(16,18)$ & \sameastraining{$c_0 = \text{Unif}(16,18)$}\\
    
    \raggedright Redshift power-law slope & $\zeta = \text{Unif}(-0.3,-0.2)$ & \sameastraining{$\zeta = \text{Unif}(-0.3,-0.2)$}\\
    
    \raggedright Peak height power-law slope & $\beta = \text{Unif}(0.55,0.85)$ & \sameastraining{$\beta = \text{Unif}(0.55,0.85)$}\\
    
    \raggedright mass--concentration power-law pivot mass ($M_\odot$) & $m_{\text{pivot,conc}} = 10^{8}$ & \sameastraining{$m_{\text{pivot,conc}} = 10^{8}$}\\
    
    Concentration dex scatter & $\sigma_{\text{conc}} = \text{Unif}(0.1,0.16)$ & \sameastraining{$\sigma_{\text{conc}} = \text{Unif}(0.1,0.16)$} \\
    
    \midrule
    \textbf{Cosmology} &  \\
    Cosmology Assumption & $\Lambda$CDM from Planck 2018 & \sameastraining{$\Lambda$CDM from Planck 2018}\\
    
    \midrule
    \textbf{Subhalos --- Section \ref{sec:subhalos}} &  \\
    \raggedright Subhalo mass function power-law index & $\gamma_{\text{sub}} \sim \text{Unif}(-1.92,-1.82)$ & \sameastraining{$\gamma_{\text{sub}} \sim \text{Unif}(-1.92,-1.82)$}\\
    
    \raggedright \textcolor{purple}{Subhalo mass function normalization $(\text{kpc}^{-2})$} & \textcolor{purple}{
    $\Sigma_{\text{sub}} \sim \mathcal{N}(\mu: 2 \times 10^{-3},\sigma: 1.1 \times 10^{-3})$} & \textcolor{purple}{$\Sigma_{\text{sub}} \sim \mathcal{N}(\mu: 2\times10^{-4}\times i,\sigma: 1.5 \times 10^{-4})^\ddag$}\\
    
    \raggedright Subhalo power-law pivot mass ($M_\odot$) &
    $m_{\text{pivot,sub}} = 10^{10}$ & \sameastraining{$m_{\text{pivot,sub}} = 10^{10}$}\\
    
    \raggedright Subhalo mass function minimum mass ($M_\odot$) &
    $m_{\text{min,sub}} = 10^{7}$ & \sameastraining{$m_{\text{min,sub}} = 10^{7}$}\\
    
    \raggedright Subhalo mass function maximum mass ($M_\odot$) &
    $m_{\text{max,sub}} = 10^{10}$ & \sameastraining{$m_{\text{max,sub}} = 10^{10}$}\\
    
    \raggedright Subhalo truncation pivot mass ($M_\odot$) &
    $m_{\text{pivot,trunc}} = 10^{7}$ & \sameastraining{$m_{\text{pivot,trunc}} = 10^{7}$}\\
    
    \raggedright Subhalo truncation pivot radius (kpc) &
    $r_{\text{pivot,trunc}} = 50$ & \sameastraining{$r_{\text{pivot,trunc}} = 50$}\\
    
    \midrule
    \textbf{LOS halos --- Section \ref{sec:los}} &  \\
    \raggedright LOS mass function normalization &
    $\delta_{\text{los}} \sim \mathcal{N}(\mu: 1.0,\sigma: 0.6)$ & \sameastraining{$\delta_{\text{los}} \sim \mathcal{N}(\mu: 1.0,\sigma: 0.6)$}\\
    
    \raggedright Mass function minimum mass ($M_\odot$) &
    $m_{\text{min,los}} = 10^{7}$ & \sameastraining{$m_{\text{min,los}} = 10^{7}$}\\
    
    \raggedright Mass function maximum mass ($M_\odot$) &
    $m_{\text{max,los}} = 10^{10}$ & \sameastraining{$m_{\text{max,los}} = 10^{10}$}\\
    
    \raggedright Minimum LOS redshift&
    $z_{\text{min,los}} = 0.01$ & \sameastraining{$z_{\text{min,los}} = 0.01$}\\
    
    \raggedright LOS redshift bin width&
    $\Delta_{z,\text{los}} = 0.01$ & \sameastraining{$\Delta_{z,\text{los}} = 0.01$}\\
    
    \raggedright LOS cone opening angle ($\arcsec$)&
    $\theta_{\text{los}} = 8.0$ & \sameastraining{$\theta_{\text{los}} = 8.0$}\\
    
    \raggedright Minimum two halo term radius (kpc)&
    $r_{\text{2halo,min}} = 0.5$ & \sameastraining{$r_{\text{2halo,min}} = 0.5$}\\
    
    \raggedright Maximum two halo term radius (kpc)&
    $r_{\text{2halo,max}} = 10.0$ & \sameastraining{$r_{\text{2halo,max}} = 10.0$}\\
    
    \raggedright Deflection angle correction redshift bin width&
    $\Delta_{z,\text{correction}} = 0.05$ & \sameastraining{$\Delta_{z,\text{correction}} = 0.05$}\\
    
    \midrule
    \raggedright \textbf{Source: COSMOS catalog --- Section \ref{sec:source}} &  \\
    
    Source redshift & $ z_{\text{source}} = 1.5$ &  \sameastraining{$z_{\text{source}} = 1.5$} \\    
    
    Maximum catalog redshift & $ z_{\text{catalog,max}} = 1.0$ & \sameastraining{$ z_{\text{catalog,max}} = 1.0$} \\
    
    Faintest catalog apparent magnitude & $ \text{mag}_{\text{faint}} = 20$ & \sameastraining{$ \text{mag}_{\text{faint}} = 20$}\\
    
    Minimum source size (pixels) & $\text{size}_\text{min,pix} = 50$ & \sameastraining{$\text{size}_\text{min,pix} = 50$} \\
    
    Minimum half-light radius (pixels) & $r_{1/2} = 10$ & \sameastraining{$r_{1/2} = 10$} \\
    
    Source rotation angle & $\phi_\text{source} \sim \text{Unif}(0,2\pi)$ & \sameastraining{$\phi_\text{source} \sim \text{Unif}(0,2\pi)$} \\
    
    x-coordinate source center ($\arcsec$) & $x_\text{source} \sim \mathcal{N}(\mu: 0, \sigma: 0.16)$ & \sameastraining{$x_\text{source} \sim \mathcal{N}(\mu: 0, \sigma: 0.16)$}\\
    
    y-coordinate source center ($\arcsec$) & $y_\text{source} \sim \mathcal{N}(\mu: 0,\sigma: 0.16)$ & \sameastraining{$y_\text{source} \sim \mathcal{N}(\mu: 0,\sigma: 0.16)$}\\
    
    Number of galaxy images & Training: 2,163 / Validation: 99 & 99 \\
    
    \bottomrule
    \end{tabularx}
    \caption{Distribution of simulation parameters in the training, validation, and test sets}
    \label{tab:dist}
    \tablecomments{For a detailed discussion of each parameter, see Section \ref{sec:simulation}. The subsets of galaxy images used for the training and test sets are disjoint from one another. In this table, $\mathcal{N}$ is the normal distribution and  $\text{Unif}$ is the uniform distribution. The subhalo mass function normalization is highlighted since it is our main parameter of interest throughout this work. For the test set distributions, any values in gray indicate that they are identical to the choices made on the training set.\\
    $\dag$: The distribution is capped to values larger than 0 \\
    $\star$: For these parameters, each test set has a mean drawn from the uniform distribution specified. \\
    $\ddag$: For test set $i$ the mean of $\Sigma_\text{sub}$ is set to $2\times10^{-4}\times i$, so for test set 4 the mean is $8\times10^{-4}$}.
\end{table*}

\subsection{Posterior Distribution for Individual Lenses}\label{sec:nn_model}

For each individual lens, our goal is to estimate the posterior of the parameters of interest given the image. To do this, we employ simulation-based inference using a neural density estimator. The estimator, $q_{F(d,\phi)}(\xi)$, approximates the posterior, $p(\xi|d,\Omega_{\text{int}})$. Here $F$ is our neural network, $q$ is a density function, $\phi$ are the parameters of that network, $d$ is the strong lensing image, $\xi$ are the physical parameters that we wish to constrain, and $\Omega_\text{int}$ is a prior distribution on the parameters $\xi$. The generation of the training set described in Section \ref{sec:sim_datasets} can be thought of as first sampling a lensing parameters from our training distribution $\xi_k \sim p(\xi|\Omega_\text{int})$ and then using our simulator, $g$, to generate an image $d_k \sim g(\xi_k)$. To train the network weights $\phi$ we then minimize the loss function:

\begin{align}\label{eq:loss_npe}
    L(\phi) &= - \sum_{k=1}^N \log q_{F(d_k,\phi)} (\xi_k).
\end{align}

In the limit where $N\to \infty$ and $q$ is sufficiently flexible, Equation \ref{eq:loss_npe} guarantees that the network will learn $q_{F(d,\phi)}(\xi) \to p(\xi|d,\Omega_{\text{int}})$ (for a proof of this statement see \citealt{papamakarios2016fast}, Appendix A). While the prior $\Omega_\text{int}$ is never explicitly enforced in our loss function, it is implicitly learned via the distribution of parameters $\xi_k$ that the network is exposed to. We have chosen to label this distribution $\Omega_\text{int}$ because it is an interim choice: it is a prior that is optimized for training but it is not necessarily the distribution we expect real lenses to follow. In Section \ref{sec:hi_inf}, we will discuss our hierarchical inference methodology for extracting the true distribution of a population of lenses.

For the model architecture, $F(d,\phi)$, we implement xResNet-34 \citep{he2016deep,he2019bag}. For the density function $q$, we use a multivariate Gaussian with a full precision matrix (see section 3.1 of \citealt{wagner2021hierarchical} for implementation details). The final fully connected layer of our xResnet-34 architecture is modified to predict the 54 parameters of our multivariate Gaussian (9 means and 45 free parameters of the precision matrix), and the first layer's filters are adapted to read monochromatic images. All layer weights are randomly initialized using the Xavier uniform initialization.

The model presented here is implemented using the \textsc{TensorFlow} \citep{tensorflow2015} library in Python and trained on a NVIDIA GeForce RTX 2080 Ti GPU. The model is first trained for 100 epochs to predict the mean and the diagonal elements of the precision matrix. Then the model is trained for a further 100 epochs with the full precision matrix, but only the parameters for the final layer of the network are updated. This training scheme helps reduce the instability generated by the precision matrix terms in the loss function. We use a batch size of 256 and the Adam optimizer. The learning rate is set to $1 \times 10^{-5}$
and the default Adam parameter values
of $\beta_1 = 0.9$, $\beta_2 = 0.999$, and $\epsilon=1\times 10^{-7}$ are used. A decay rate of 0.98 is applied to the learning rate at the end of each epoch. In total, training takes $\approx 96$ hours.

Each individual image is normalized to have a standard deviation of 1, and the network is trained to predict output parameters that are normalized to have mean 0 and standard deviation 1 on the full training set. The constants used for this normalization are saved so that the network outputs can be translated back to the physical parameters. As an additional training augmentation, each batch of images is randomly rotated before being fed into the network, and the lensing parameters are corrected accordingly. At inference time, we find that repeating this same random rotation improves the quality of inference. Therefore, all of the mean predictions, except those for $x_\text{lens},y_\text{lens}$, are averaged over 100 random rotations of the input image. The covariance predictions are not averaged over rotations.

\subsection{Hierarchical Inference}\label{sec:hi_inf}

The final goal of our analysis is to constrain the distribution governing the normalization of the subhalo mass function, $\Sigma_\text{sub}$, given a set of strong lensing images. We can write this as $p(\Omega|\{d\})$, where $\Omega$ is the population level distribution for our lensing parameters $\{\xi\}$, and $\{d\}$ is our set of strong lensing images. As we introduced in Section \ref{sec:nn_model}, the model does not approximate the likelihood $p(\xi|d)$ but rather the posterior $p(\xi|d,\Omega_\text{int})$. We must chose $\Omega_\text{int}$ at training time, and therefore we cannot vary it to calculate $p(\Omega|\{d\})$. Instead we must re-weight our network's posterior estimates, $q_{F(d,\phi)} (\xi)$, by the ratio of the likelihood of drawing a parameter $\xi$ given the proposed distribution $\Omega$ versus the training distribution $\Omega_\text{int}$:

\begin{align}
    p(\Omega | \{ d \})&=  \!\begin{aligned}[t] \underbrace{p(\Omega)}_\text{$\Omega$ prior} \times & \underbrace{\prod_k^{N_\text{lens}} \frac{p(d_k|\Omega_\text{int})}{p(\{d\})}}_\text{normalizing factor} \times \\
    & \underbrace{\prod_k^{N_\text{lens}} \int \frac{p(\xi|\Omega)}{p(\xi|\Omega_\text{int})} q_{F(d_k,\phi)} (\xi) \ d \xi}_\text{importance-sampling integral}. \end{aligned}
     \label{eq:post_omega}
\end{align}

A more detailed derivation of Equation \ref{eq:post_omega} can be found in appendix C of \cite{wagner2021hierarchical}. Note that only the first and third terms of the right-hand-side depend on $\Omega$. The first term is simply a hyperprior on the distribution $\Omega$\footnote{This hyperprior allows for us to enforce physical constraints on the proposed distribution, including constraining the inferred population mean of the SHMF normalization to be greater than 0.}, the second term is a constant normalizing factor, and the third term encodes all of the constraining power provided by the population of images. The ratio of $p(\xi|\Omega)$ to $p(\xi|\Omega_\text{int})$ is often called an importance-sampling weighting. To keep the integral tractable, we must chose $\Omega_\text{int}$ such that the term $p(\xi|\Omega)/p(\xi|\Omega_\text{int})$ is finite for all $\xi$. Practically speaking, this means that the training distribution $\Omega_\text{int}$ must be broader than any distribution $\Omega$ we would like to infer. Similarly, while it is possible to evaluate the integral numerically (see \citealt{wagner2021hierarchical}), having an analytic solution for the integral allows for faster sampling of $p(\Omega|\{d\})$. These two constraints inform our choice of training distribution (broad, Gaussian distribution) and test distributions (narrow, Gaussian distributions).

We use Equation \ref{eq:post_omega} to place constraints on the distribution of $\Sigma_\text{sub}$. To do this, we first pass the lens images in our test set through our trained network. This gives us the posterior prediction $q_{F(d_k,\phi_j)}$ for each lens $k$ and network $j$. We then sample Equation \ref{eq:post_omega} to get the posterior on $\Omega$. For the experiments conducted in this work, both $\Omega$ and $\Omega_\text{int}$ are described by nine means and nine standard deviations. These means and standard deviations describe the distribution of our eight main deflector parameters and the subhalo mass function normalization, $\Sigma_\text{sub}$. To sample the posterior we use an ensemble sampler with affine invariance \citep{goodman2010ensemble} implemented through the \textsc{emcee} package\footnote{\url{https://emcee.readthedocs.io}} \citep{emcee}. 

\section{Results}\label{sec:results}

\begin{figure*}
    \centering
    \includegraphics[scale=0.5]{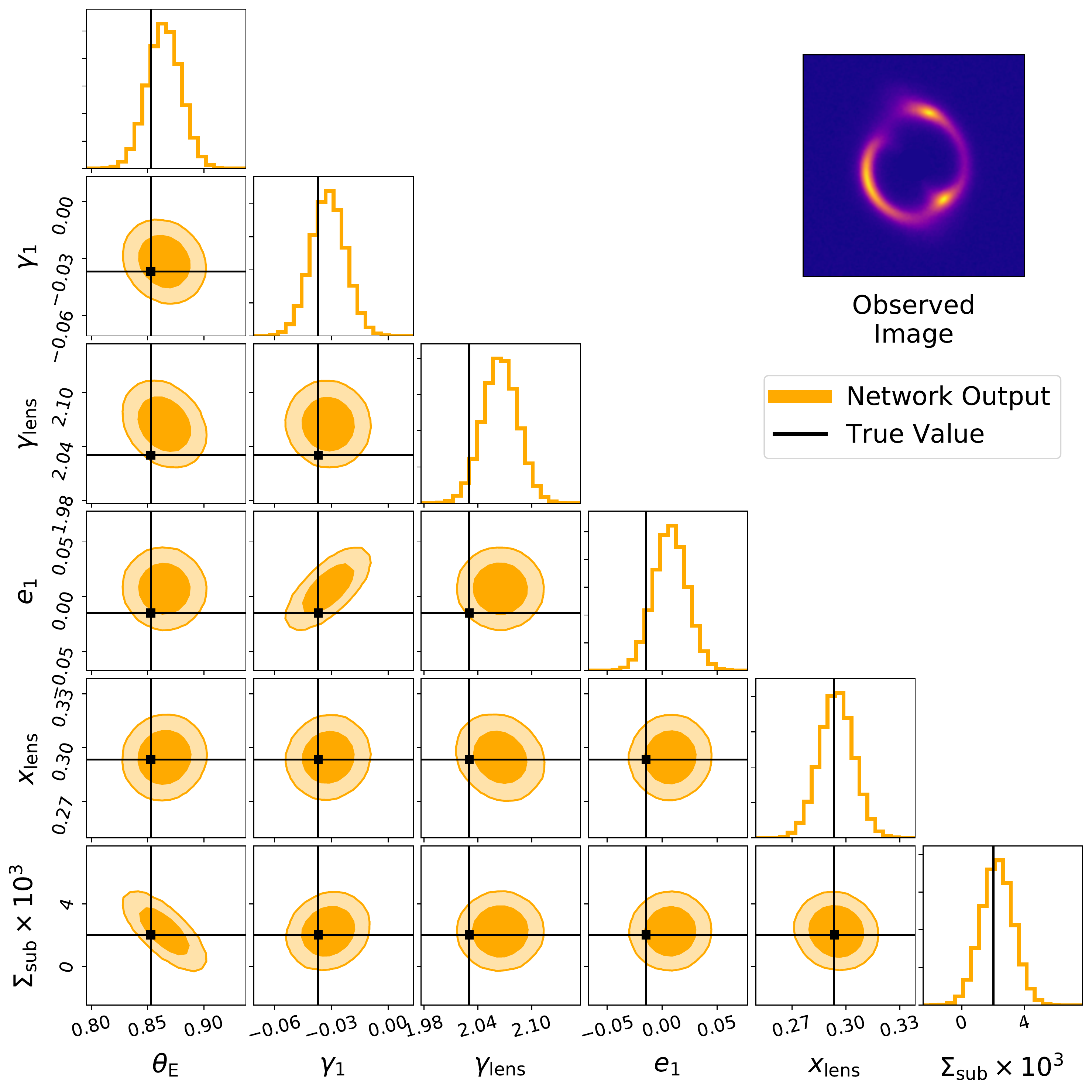}
    \caption{Example of the estimated posteriors output by our network for a simulated observation. The black points, labeled `true value,' represent the input parameters used to simulate the image in the upper-right. The yellow contours represent the multivariate Gaussian output by our neural posterior estimator. The darker and lighter contours correspond to the 68\% and 95\% confidence intervals respectively. The posterior output for $\gamma_2,e_2$, and $y_\text{lens}$ has been omitted to avoid visual clutter.}
    \label{fig:corner_plot}
\end{figure*}

\begin{figure*}
    \centering
    \includegraphics[scale=0.35]{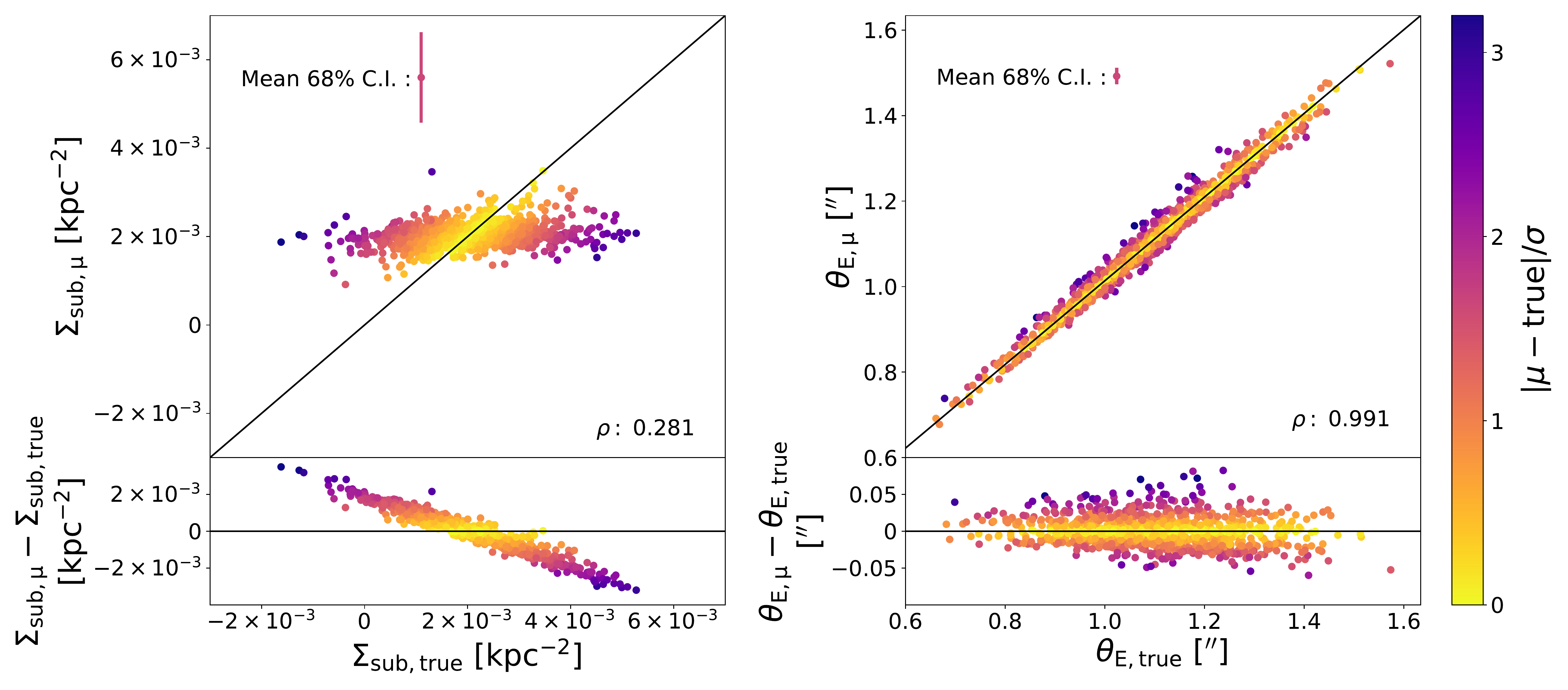}
    \caption{Comparison of the network predicted means and the true input values for all of the lenses in our validation set. The left plot shows the comparison for the SHMF normalization $\Sigma_\text{sub}$, and the right plot shows the comparison for the Einstein radius $\theta_{E}$. For both plots, one point corresponds to the predictions of one lens. The color of the point is defined by the distance between the predicted mean and the truth in units of the predicted standard deviation. The colorbar for both plots is on the far right. The upper left includes a visualization of the average standard deviation being predicted by the network (equivalent to the 68\% confidence interval). The $\rho$ value printed in the plots represents the Pearson correlation coefficient between the predicted mean and the truth. The remaining parameters are shown in Figure \ref{fig:val_remain}}
    \label{fig:confidence_plot}
\end{figure*}

\subsection{Performance on Individual Lenses}\label{sec:perf_train}

As an initial test of the performance of our neural posterior estimator, we run the network on our validation set. As discussed in Section \ref{sec:sim_datasets}, our validation set draws from the same underlying parameter distributions used for the training set, but uses a held-out set of 99 galaxy sources. In Figure \ref{fig:corner_plot} we show the posterior predictions for one randomly selected validation image. The most notable feature is that the posteriors on $\Sigma_\text{sub}$ are nearly as wide as the training distribution. The neural posterior estimator predicts a standard deviation of $1.0\times10^{-3} \ \text{kpc}^{-2}$ for the SHMF normalization, $\Sigma_\text{sub}$, compared to the standard deviation of $1.1 \times 10^{-3} \ \text{kpc}^{-2}$ used to draw the SHMF normalization in the training dataset. The network posterior also includes a strong covariance between the SHMF normalization and the Einstein radius, $\theta_\text{E}$. This covariance is a product of our definition of the Einstein radius: as we add more subhalos to our simulation, we increase the effective mass of our main deflector and therefore increase the observed radius of the ring in the image. However, the `true' value of the Einstein radius only accounts for the mass in our PEMD main deflector. Therefore, simultaneously decreasing the Einstein radius of the main deflector and increasing the mass in subhalos produces a similar observation. The negative correlation output by our network is a consequence of that degeneracy \footnote{Due to the strength of this degeneracy, we include additional tests in Appendix \ref{app:einst_radius} that confirm that the inferred $\Sigma_\text{sub}$ for a test population of lenses is not systematically biased by shifts in the underlying distribution of $\theta_\text{E}$.}.

Figure \ref{fig:confidence_plot} compares the predicted mean values of SHMF normalization and the Einstein radius to the true value across the entire validation dataset. The Einstein radius appears to be well constrained by the data, with the predicted mean being tightly correlated to the true value ($\rho = 0.991$)\footnote{$\rho$ is the Pearson correlation coefficient.}. While not visualized in Figure \ref{fig:confidence_plot}, the same is true for all eight of the main deflector parameters our network predicts (see Appendix \ref{app:val_remain}). The SHMF normalization, however, is rather poorly constrained, with the mean predictions clustering around the training set mean of $2.0\times10^{-3} \ \text{kpc}^{-2}$ and weakly correlated to the true value ($\rho = 0.281$). In agreement with previous work \citep{perreault2017uncertainties,wagner2021hierarchical,pearson2021strong}, the network is capable of precisely and accurately constraining the main deflector parameters. But the estimated posteriors for the SHMF normalizations are dominated by the training prior, with little information extracted from any individual lens. The signal produced by a fixed SHMF normalization is highly stochastic (see Appendix \ref{app:res_an} for examples), so the comparatively weak correlation produced by our network likely reflects the poor information content of the data. 

However, as we will explore in Section~\ref{sec:recon_shmf}, a weakly correlated but statistically consistent posterior can be sufficient to extract population constraints. Figure~\ref{fig:confidence_plot} colors the predicted means according to their distance from the truth in units of the predicted standard deviation. Because the one-dimensional predictions are Gaussian, a statistically consistent posterior would have the true values Gaussian distributed around the mean. That is what we find on the validation set: $69\%,95\%,$ and $99\%$ of the true values fall within one, two, and three standard deviations of the mean respectively. 

\subsection{Reconstructing the Subhalo Mass Function}\label{sec:recon_shmf}

\begin{figure*}[t!]
    \centering
    \includegraphics[scale=0.35]{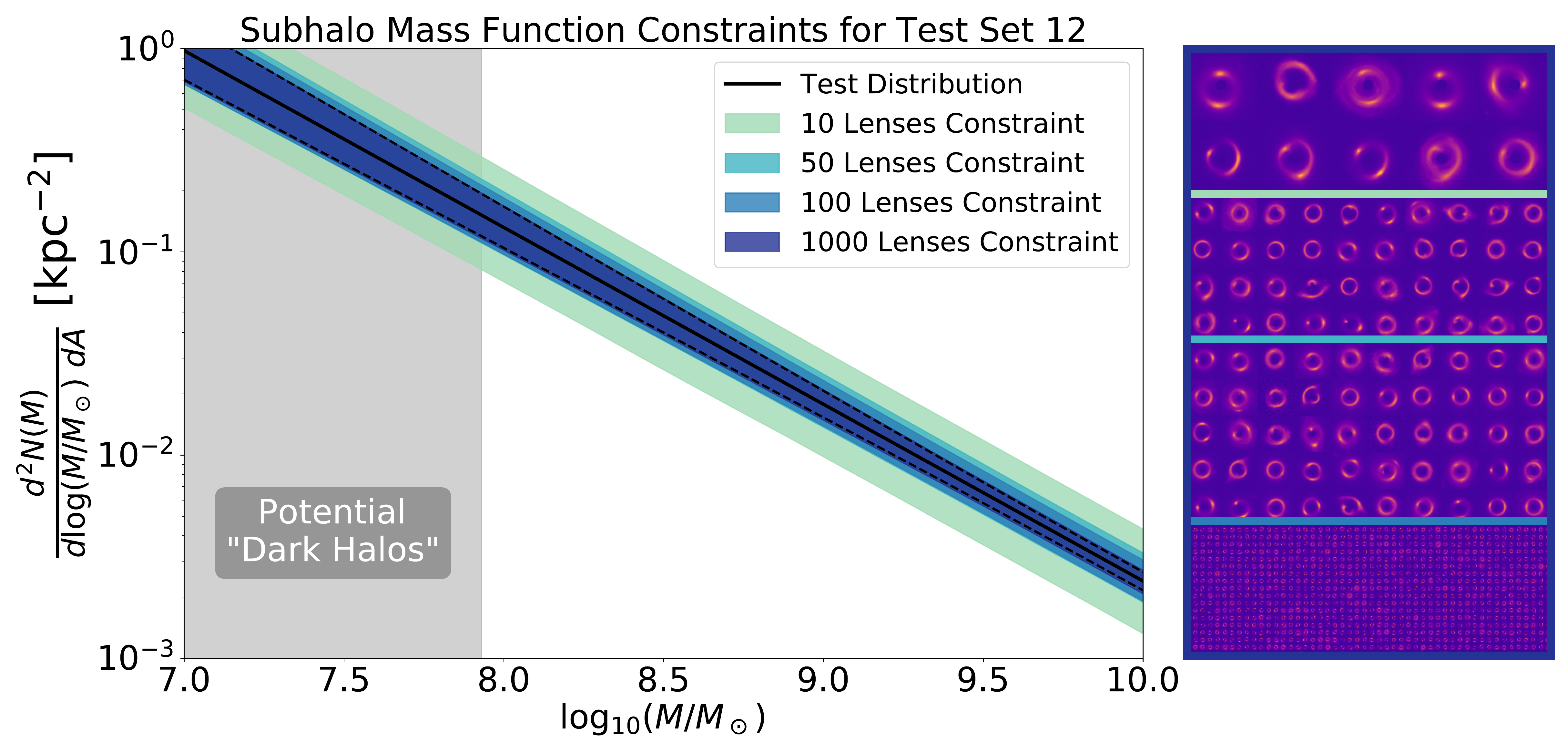}
    \caption{The subhalo mass function constraints for test set 12 as a function of the number of lenses used in the analysis. In the left hand plot, the black solid and dashed lines show the range containing 95\% of the SHMFs coming from the test distribution. The colored contours shows the hierarchical inference estimate, which includes marginalizing over the uncertainty in the mean and standard deviation of $\Sigma_\text{sub}$. The gray contour labeled `potential dark halos' shows the current limits on halos without luminous counterparts from Milky Way satellites \citep{nadler2020milky}. The right hand plot shows the lenses included in the 10, 50, 100, and 1000 lens analysis, with the colors corresponding to the contours on the left hand plot.}
    \label{fig:shmf_12}
\end{figure*}

\begin{figure*}
    \centering
    \includegraphics[scale=0.35]{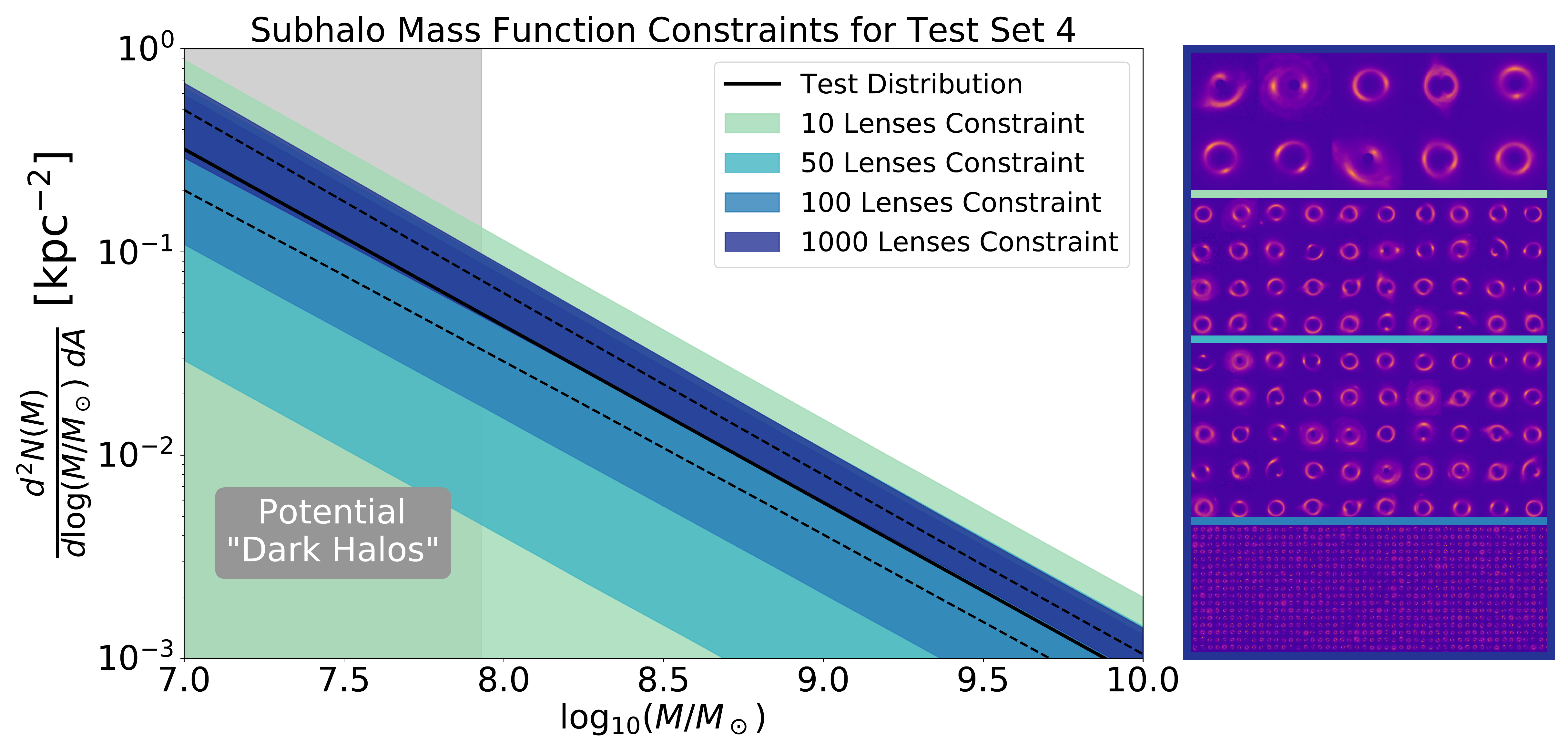}
    \caption{The same subhalo mass function constraints as in Figure \ref{fig:shmf_12}, but now for test set 4.}
    \label{fig:shmf_4}
\end{figure*}

\begin{figure*}
    \centering
    \includegraphics[scale=0.45]{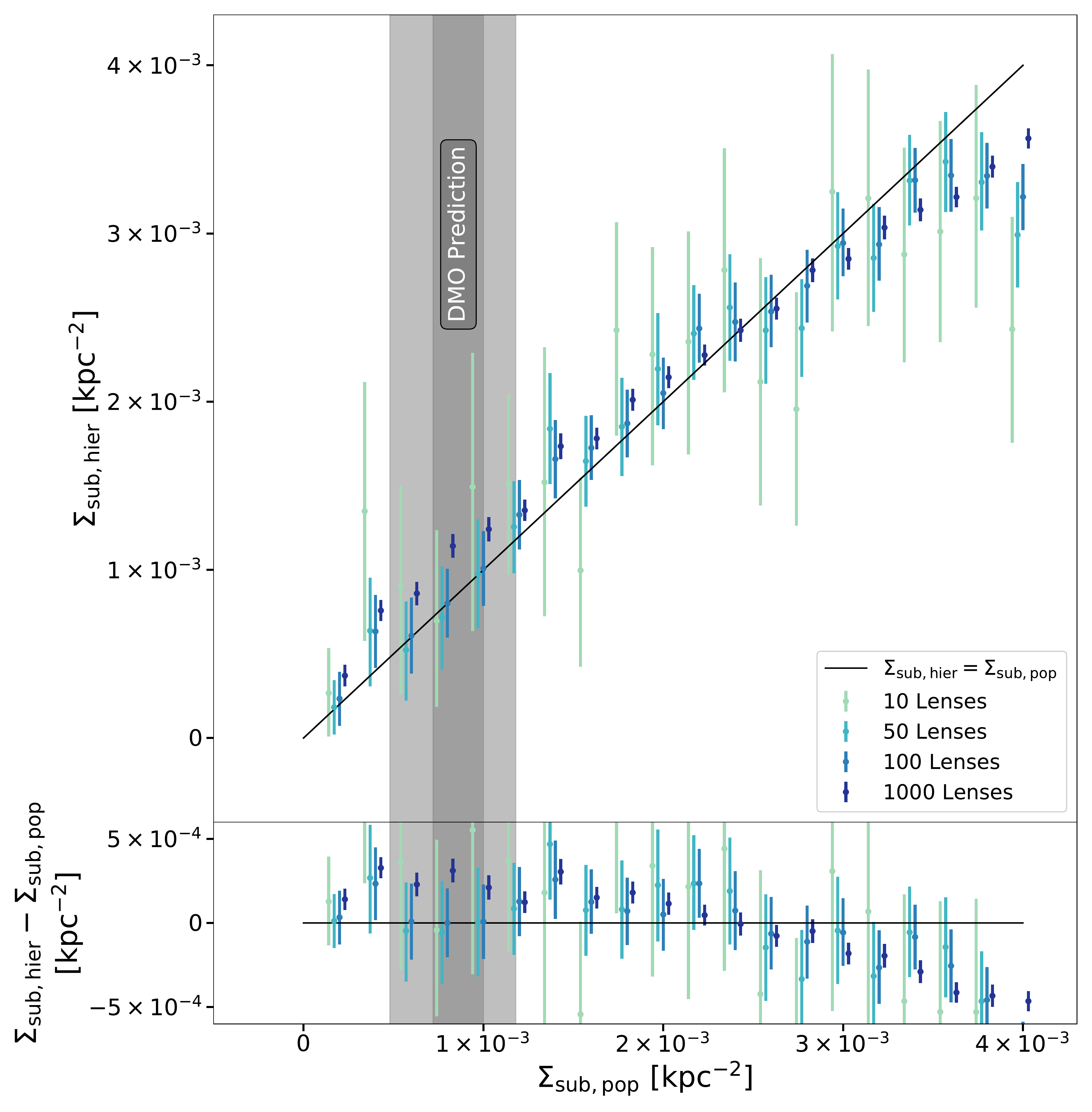}
    \caption{Inferred ($\Sigma_\text{sub,hier}$) versus true  ($\Sigma_\text{sub,pop}$) mean of the SHMF normalization distribution (top), and the difference between these quantities (bottom). Each set of bars represents constraints using 10, 50, 100, or 1000 lenses from one of our twenty test sets. For a given test set, all analyses have the same $\Sigma_\text{sub,pop}$, but the coordinates of the plotted points have been slightly offset for visual clarity. The gray shaded contours contain 68\% and 100\% of the SHMF normalizations for $\approx 10^{13} M_\odot$ host halos at redshift $z=0.5$ measured on a suite of high-resolution cosmological dark matter-only (DMO) zoom-in simulations (E. Nadler et al. 2022, in preparation). The error bars on the inferred $\Sigma_\text{sub,hier}$ show the 68\% interval derived from the hierarchical inference pipeline.}
    \label{fig:response_plot}
\end{figure*}

To test the network's ability to constrain a population of lenses, we run our hierarchical inference pipeline on the twenty test sets described in Section \ref{sec:sim_datasets}. We first pass the 1000 images in each of our test sets through our neural posterior estimator. Then, we use equation \ref{eq:post_omega} to simultaneously constrain the population distribution of the SHMF normalization and the eight main deflector parameters. Figure \ref{fig:shmf_12} and Figure \ref{fig:shmf_4} show the inferred SHMF distribution for test sets 12, $(\Sigma_\text{sub,pop} = 2.4 \times 10^{-3})$, and 4, $(\Sigma_\text{sub,pop} = 0.8 \times 10^{-3})$, respectively. The constraints on the SHMF include the uncertainty on the inferred population mean, $\Sigma_{\text{sub},\mu}$, and standard deviation, $\Sigma_{\text{sub},\sigma}$, of the SHMF normalization. The constraining power at very low subhalo mass ($<10^{8} M_\odot$) is driven by our prior assumption about the SHMF slope, $\gamma_\text{sub}$. For test set 12, 10 lenses are sufficient to strongly disfavor the existence of halos without subhalos. As we push the analysis from 10 to 1000 lenses, the inferred SHMF converges toward the true test distribution, with the uncertainty in the mean and scatter in the 1000 lens analysis being subdominant to the intrinsic variation in the lens population. For test set 4, the 10 lens analysis fails to exclude the existence of halos without subhalos within its 95\% confidence interval, but the 50 lens analysis already strongly disfavors that scenario. Unlike test set 12, we can see an upward bias in the inferred SHMF for the 1000 lens analysis, but the 10, 50, and 100 lens analyses are consistent with the truth.

In Figure \ref{fig:response_plot} we show the constraints on the population mean of each test set as a function of the true population mean. Despite the weak correlation discussed in Section \ref{sec:perf_train}, the analysis pipeline returns a nearly linear response between the true population means and the inferred population means. For the 10, 50, and 100 lens analyses we find that the pipeline returns unbiased results for all but the largest values of $\Sigma_\text{sub,pop}$. For $\Sigma_\text{sub,pop}>3.6 \times 10^{-3} \ \text{kpc}^{-2}$, the inferred population mean for 10, 50, and 100 lenses is an underestimate. When we push the analysis to 1000 lenses, the inferred mean is still linearly determined by the truth, but it now exhibits a systematic bias for nearly all values of $\Sigma_\text{sub,pop}$. Specifically, it pushes its estimates toward the training prior, overestimating values of $\Sigma_\text{sub,pop}$ less than $2 \times 10^{-3}$ and underestimating values of $\Sigma_\text{sub,pop}$ greater than $2 \times 10^{-3}$. This pattern of bias suggests that the network has an imperfect understanding of the training prior $\Omega_\text{int}$. In order to map from the prior-dominated constraints shown in Figure~\ref{fig:confidence_plot} to the hierarchical results in Figure \ref{fig:response_plot}, we divide by the interim prior as shown in Equation \ref{eq:post_omega}. This division should cancel the influence of the training prior in the network's posterior. However, the network learns the training prior implicitly from the training dataset. If the network's learned estimate of the training prior is too wide, the division will under-correct the network posteriors and bias the overall estimate toward the training prior mean. This is consistent with the shift we see in Figure \ref{fig:response_plot} for the 1000 lens analysis. The discrepancy in the learned and true training prior is sufficiently small that it does not appear to meaningfully impact the 10, 50, or 100 lens analyses.

Figure \ref{fig:response_plot} also includes the range of SHMF normalizations for $\approx 10^{13} M_\odot$ halos found in a suite of high-resolution cosmological dark matter-only (DMO) zoom-in simulations (E. Nadler et al. 2022, in preparation). The SHMF normalization has been measured at a redshift of $z=0.5$ and includes the subhalos within the projected virial radius with masses in the range $[9.4 \times 10^{8} M_\odot,1 \times 10^{10} M_\odot]$. We plot the region containing 68\% and 100\% of the SHMF normalizations. The `host-to-host' scatter in the measured SHMF normalization is connected to the secondary properties of the main deflector halo and its assembly history (See \citealt{mao2018beyond} for a review). Distinguishing between a mean SHMF normalization in the upper and lower halves of this range would enable meaningful measurements of the concentration of main deflector halos and the environment in which they form. Alternatively, if a strong lensing analysis returned a $\Sigma_\text{sub,hier}$ constraint that did not overlap with this region, then there would exist a tension between strong lensing measurements and DMO predictions that could not be explained by selection effects. The 50, 100, and 1000 lens analyses produce constraints smaller than the DMO predicted host-to-host variation in the SHMF normalization, and even the 10 lens analysis can detect a tension between strong lensing measurements and DMO predictions for sufficiently high $\Sigma_\text{sub,pop}$. In terms of detecting the existence of subhalos (i.e.\ inferring $\Sigma_\text{sub,hier} > 0$), 50 lenses are sufficient for the full range of SHMF normalizations present in the DMO simulation suite. The 50, 100, and 1000 lens analyses provide a $32\%$, $22\%$, and $6\%$ precision measurement of the normalization in the region favored by the DMO simulation suite. 

\section{Discussion}\label{sec:discuss}

We have presented a pipeline for inferring the SHMF normalization of a large population of lenses and probed the robustness of that pipeline on test datasets with a wide range of SHMF normalizations and main deflector parameters. In this section we will discuss the limitations of our analysis, ways in which these limitations can be addressed, and avenues for future work.

Despite weak, prior-dominated constraints on the level of individual lenses, analyzing the neural posterior estimator output hierarchically allows us to accurately reconstruct the SHMF normalization across a wide portion of parameter space. This includes inference on datasets that are built with COSMOS sources not seen during training and marginalizing over the effects of line-of-sight halos and the mass--concentration relation. The results also scale well to large datasets, as adding more lenses to our analysis consistently improves the constraining power of our network and has a negligible impact on the computational time of the pipeline. However, for the 1000 lens analysis and the high SHMF normalization region, the normalizations inferred by the pipeline show systematic bias. Therefore, with the current choices, the analysis we present is limited to lens populations with $\approx 100$ lenses or inferred SHMF normalizations that are well-sampled by the training set. There are a number ways to potentially circumvent this limitation. For example we could conduct sequential neural posterior estimation (SNPE), a scheme of iterative retraining where the inferred SHMF population is used as a prior for a new training set (see \citealt{greenberg2019automatic} for a review of SNPE implementations). By construction, this additional training would ensure that the inferred SHMF normalizations fall in a region well sampled by the training distribution, reducing the tension between the test population and the assumptions made during training. We leave introducing retraining to our methodology for future work.

Additionally, while we have made significant efforts to create realistic and complex simulations, there are still assumptions in our current analysis that will need to be modified for observed lenses. For example, the use of COSMOS galaxies for our sources enables us to marginalize over realistic morphological source complexity but requires cutting on nearby, well-resolved galaxies. This selection assumes that low redshift galaxies are representative of high redshift galaxies. The population level differences in the morphology of high redshift and low redshift galaxies may be degenerate with the small-scale subhalo perturbations and could generate a bias on the inferred SHMF. One approach would be to develop more realistic simulated images of high redshift source galaxies, that can be realized at the required super-resolution and input into our simulation pipeline.   

We have also assumed a fixed cosmology throughout our analysis. Modifying the cosmological parameters would affect our halo profiles, the volume from which we sample line-of-sight halos, and the distances that set the angular size of the galaxies and halos in our simulation. Similarly, we have assumed broad priors on the mass--concentration parameters, the slope of the SHMF, and the line-of-sight halo mass function\footnote{In Appendix \ref{app:delta_los} we explore how changing the distribution of line-of-sight halo normalizations affects the inferred subhalo mass function normalization.}. This allows us to probe the sensitivity of our analysis when it is marginalized over the existing theoretical uncertainties, but a broad, mis-centered prior can systematically shift the inferred SHMF normalization. On real data, we could choose a carefully constrained set of theoretical priors and recognize that the analysis is conditioned on those assumptions. Alternatively, the parameters governing the source population, SHMF slope, mass--concentration parameters, and line-of-sight halos can be fit on a lens-by-lens level as we do for the main deflector parameters. We could then infer these parameters hierarchically from the data. The multivariate Gaussian posterior we use is not sufficiently expressive for this task, but a network leveraging normalizing flows or another more flexible posterior could circumvent this limitation. For the parameters on which we assume broad priors, either more constrained priors or a hierarchical inference approach would yield tighter constraints than the marginalization we use in this work. In that respect, the sensitivity to the SHMF normalization discussed in Section \ref{sec:recon_shmf} is conservative. We leave exploring this extension of our hierarchical approach for future work.

Lastly, while our simulation and inference pipelines account for the full complexity of low-mass halos, we have focused our analysis on the normalization of the SHMF. Similarly, we have only considered the use of HST images. These choices allow us to illustrate our ability to infer dark matter parameters from large samples of lenses but leave two major avenues for future work: exploring the use of different imaging datasets and quantifying low-mass halo parameters beyond the SHMF normalization. In terms of datasets, interferometers like the Atacama Large Millimeter/submillimeter Array are capable of returning higher resolution images of strong lenses, albeit with the addition of significant data complexity \citep{hezaveh2013alma}. No part of our methodology is inherently resolution limited. Incorporating interferometry measurements would require either modifying the observational effects to include the artifacts introduced from Fourier space measurements or modifying the network architecture to take the Fourier space signal as input. In turn, these higher resolution images could yield substantially more constraining power on the SHMF. In terms of the parameters we measure, future studies will also want to constrain a low-mass cut-off for both the SHMF and line-of-sight mass functions. This cut-off captures the behavior expected in several alternative dark matter models, and would allow us to better quantify what mass range our model is sensitive to as a function of the number of lenses in our analysis\footnote{For some preliminary discussion of our sensitivity to the mass cut-off see Appendix \ref{app:res_an}}.

\section{Conclusions}\label{sec:conc}

We present a simulation-based inference methodology for measuring the SHMF normalization of strong lensing systems. Leveraging our simulation package, \textsc{paltas}, we have trained a neural posterior estimator on a set of 500,000 synthetic lens systems with sources pulled directly from the COSMOS field, realistic low-mass halo distributions, and HST observational effects. We have tested this network against a series of test datasets and have demonstrated its ability to generalize to COSMOS galaxies not seen in the training dataset. Our network returns SHMF normalization constraints that are dominated by the prior on the individual lens level, but our hierarchical inference can extract unbiased constraints for populations with 10, 50, and 100 lenses across a wide range of normalizations, including the SHMF normalizations predicted by DMO simulations. Analyses with 1000 lenses are computationally accessible and improve the precision at the cost of some systematic bias. We discuss how iterative retraining of the network using simulated datasets that match the inferred population would likely alleviate this bias.

Galaxy--galaxy strong gravitational lenses are sensitive to dark matter on the small scales that best constrain alternatives to CDM. Previous work with strong lenses, including in the subhalo context, has shown that neural density estimators can extract accurate and precise parameter estimates and scale to populations with thousands of images. Extending these analysis techniques to observed lenses requires simulation tools capable of capturing the complexity of real data as well as rigorous testing of the robustness of the networks. We believe that this work makes significant contributions to both of these challenges: our simulated datasets have been carefully constructed to be realistic and are produced with a publicly available and well-documented software package. The combined neural posterior estimator and hierarchical inference method we propose scales to large lens populations and has been tested for robustness against unseen COSMOS galaxies, shifts in the main deflector parameter distribution, and variations in the underlying SHMF. We have shown that our method performs well across nearly all of these robustness tests, and have discussed improvements that can be made to future analyses to address the shortcomings that do exist. We are confident that the simulation-based inference pipeline presented in this work is capable of constraining dark matter substructure using both the HST strong lensing images that exist today and the datasets that will become available with the next generation of wide-field optical imaging surveys.

\subsection*{Acknowledgments}

We would like to thank Phil Mansfield for numerous helpful discussions during the development of this work. We would also like to thank María Luz Raggio for helping with the design of the \textsc{paltas} logo. SWC was supported by NSF Award DGE-1656518 and a Stanford Data Science Fellowship.

This work received support from the Kavli Institute for Particle Astrophysics and Cosmology through Kavli Fellowships to SB and JA and the U.S. Department of Energy under contract number DE-AC02-76SF00515 to SLAC National Accelerator Laboratory.

\bibliography{main}

\appendix

\section{Subhalo Profiles} \label{app:subhalo}

The subhalos in our simulation are modeled as a truncated Navarro-Frenk-White (NFW) radial density profile \citep{baltz2009analytic}:

\begin{align}
    \rho_\text{tNFW}(r) = \frac{\rho_\text{sub}}{\frac{r}{r_{s,\text{sub}}}\left( 1 + \frac{r}{r_{s,\text{sub}}}\right)^2}\frac{r_t^2}{r^2+r_t^2},
\end{align}
where $\rho_\text{sub}$ is the amplitude of the NFW density function in units of $M_\odot/kpc^3$, $r$ is the radial position in units of kpc, $r_{s,\text{sub}}$ is the scale radius in units of kpc, and $r_t$ is the truncation radius in units of kpc. Both $r_{s,\text{sub}}$ and $\rho_\text{sub}$ are calculated from the mass, $m_\text{sub}$, and concentration, $c_\text{sub}$, of the subhalo. The mass is drawn from Equation \ref{eq:shmf}, and the concentration is drawn from the mass--concentration relation presented in \cite{gilman2020constraints}:

\begin{align}\label{eq:mass_concentration}
    c_\text{sub}(m,z) &= c_0 (1+z)^\zeta \left( \frac{\nu(r_\text{peak}(m_\text{sub}),z_\text{sub})}{\nu(r_\text{peak}(m_\text{pivot,conc}),0)} \right)^{-\beta},
\end{align}
where $c_0$ is the normalization of the mass--concentration relation, $\zeta$ is the redshift power-law slope, $\beta$ is the peak height power-law slope, $m_\text{pivot,conc}$ is the mass--concentration pivot mass. The function $\nu$ is the peak height function (\citealt{doroshkevich1970spatial,peebles1980large}, see \citealt{mo2010galaxy} for the equations) at the subhalo redshift $z_\text{sub}$ for peak radius $r_\text{peak}(m_\text{sub})$ defined by:

\begin{align}
    r_\text{peak}(m_\text{sub}) = \left( \frac{3m_\text{sub}}{4 \pi \rho_{m,0}} \right)^{1/3}. 
\end{align}
Here $\rho_{m,0}$ is the matter density at redshift zero. To calculate the peak height we use the power spectrum derived from the \cite{eisenstein1998baryonic} transfer function. In line with the literature \citep{dutton2014cold,diemer2015universal,diemer2019accurate}, we add an additional scatter, $\sigma_{\text conc}$, to Equation \ref{eq:mass_concentration}. Given the mass and concentration, the scale radius and amplitude of the density function are given by:

\begin{align}
    r_{s,\text{sub}} &= \frac{1}{c}\left(\frac{3 \ m_\text{sub}}{4 \pi (200\rho_\text{crit}(z_\text{sub}))}\right)^{1/3} \\
    \rho_\text{sub} &= \frac{m_\text{sub}}{4 \pi r_{s,\text{sub}}^3 \left( \log(1+c) - \frac{c}{1+c} \right)},
\end{align}
where $\rho_\text{crit}(z)$ is the critical density of the universe at the redshift $z_\text{sub}$ of the subhalo. 
The truncation radius is determined by the subhalo's position in the host using:

\begin{align}
    r_t = 1.4 \left(\frac{m_\text{sub}}{m_\text{pivot,trunc}}\right)^{1/3} \left(\frac{r_\text{sub}}{r_\text{pivot,trunc}}\right)^{2/3},
\end{align}
where $m_\text{pivot,trunc}$ is the truncation pivot mass in units of $M_\odot$, $r_\text{sub}$ is the radial distance of the subhalo from the host center in units of kpc, and $r_\text{pivot,trunc}$ is the truncation pivot radius in units of kpc. 

\section{Line-of-Sight Mass Function}\label{app:los}

We make two modifications to the Sheth--Tormen halo mass function. The first is a scaling parameter, $\delta_\text{los}$, that accounts for uncertainties in overall normalization of the line-of-sight mass function. We also include a contribution from the two-point halo correlation function $\xi_\text{2 halo}(r,m_\text{host},z_\text{host})$. This term accounts for the overdensity of halos relative to the mean matter density of the universe near a massive halo. On large scales, we can describe the two-point halo correlation correlation function as:

\begin{align}
    \xi_\text{2 halo}(r,m_\text{host},z_\text{host}) &= b(m_\text{host},z) \xi_\text{lin}(r,z_\text{host}).
\end{align}
Here, $b(m_\text{host},z)$ is a mass-dependent linear halo bias parameter (\citealt{kaiser1984spatial,bardeen1986statistics,mo1996analytic,jing1998accurate}; see \citealt{desjacques2018large} for a review) that must be included because dark matter halos are biased tracers of the underlying distribution of matter. The dependence of the bias parameter on the host mass takes into account high-mass halos being comparatively more likely in overdense regions and low-mass halos being comparatively more likely in underdense regions. In this work, we use the bias model presented in \cite{tinker2010large}. The remaining term, $\xi_\text{lin}(r,z_\text{host})$, represents the linear matter-matter correlation function at redshift $z_\text{host}$. We use the correlation function derived from the \cite{eisenstein1998baryonic} transfer function. This correlation function contribution is included for halos within the range $[r_{\text{2halo,min}},r_{\text{2halo,max}}]$ of the host. Recent work has argued that lensing can produce clustering that exceeds the two-point halo correlation \citep{lazar2021out}, but we do not model that additional signal here.

With everything included, the line-of-sight halo mass function is given by the equation:

\begin{align}
    \frac{d^2 N_\text{los}}{dV \ dm_\text{los}} = \delta_\text{los}(1+\xi_\text{2 halo}(r,m_\text{host},z_\text{host})) \left[\frac{d^2 N_\text{los}}{dV \ dm_\text{los}}\right]_\text{ST},
\end{align}
where the Sheth--Tormen halo mass function is given by:

\begin{align}\label{eq:st}
    \left[\frac{d^2 N_\text{los}}{dV \ dm_\text{los}}\right]_\text{ST} &= -\frac{1}{3}\frac{\nu f(\nu)}{m^2} \frac{d \log \sigma(r_\text{peak})}{d \log r_\text{peak}} \rho_m.
\end{align}
The definitions of the radius $r_\text{peak}$ is given in Appendix \ref{app:subhalo}, with the dependence of on $m_\text{los}$ and $z_\text{los}$ left implicit in Equation \ref{eq:st} for conciseness. The peak height, $\nu$, is related to the rms variance of the linear density field $\sigma(r_\text{peak})$ by:

\begin{align}
    \nu(r_\text{peak}(m_\text{los}),z_\text{los}) = \frac{\delta_c(z_\text{los})}{\sigma(r_\text{peak}(m_\text{los}),z_\text{los})},
\end{align}
with $\delta_c(z_\text{los})$ being the linear overdensity threshold for halo collapse at redshift $z_\text{los}$ (see \citealt{mo2010galaxy} for a derivation). The functional form of $\nu f(\nu)$ is given in \cite{sheth2001ellipsoidal}:

\begin{align}
    \nu f(\nu) &= 2 A_\text{ST} \left(1+\frac{1}{\nu'^{2q_\text{ST}}}\right) \frac{\nu'}{(2\pi)^{1/2}} \exp \left(-\frac{\nu'^2}{2} \right) \\
    \nu' &= \sqrt{a_\text{ST}} \nu,
\end{align}
with $A_\text{ST}=0.32218$, $q_\text{ST}= 0.3$, and $a_\text{ST}=0.707$. The Sheth--Tormen halo mass function is non-trivial to draw from, however for a relatively small mass range it can be well-approximated by a power-law. Therefore, in practice, we draw from a power-law with normalization and slope set by minimizing the log squared distance to the Sheth--Tormen halo mass function in the range $[m_\text{min,los},m_\text{max,los}]$. The parameters $m_\text{min,los}$ and $m_\text{max,los}$ also set the minimum and maximum line-of-sight halo mass that will be rendered.

\section{COSMOS Images}\label{app:source}

A more detailed summary of the GREAT3 dataset we use can be found in \cite{mandelbaum2014third}. The important points for our use are:  

\begin{itemize}
    \item The COSMOS images have been processed using \textsc{MultiDrizzle} (see \citealt{fruchter2002drizzle} for a summary of the drizzle algorithm and \citealt{gonzaga2012drizzlepac} for a discussion of a modern implementation of the full pipeline). This pipeline deals with the geometric distortion, sky subtraction, and cosmic ray rejection. As part of this process, it combines several exposures dithered at sub-pixel intervals. It is therefore capable of returning a smaller pixel scale in the final co-added images, which for the COSMOS patch is set to 0.03\arcsec.
    \item Within the COSMOS patch, the sources are selected using the strategy outlined in \cite{leauthaud2007weak}. On top of the selection cuts described there, there are additional selection cuts made to reject non-galaxy objects, galaxies with imaging defects, and galaxies that do not have a reliable photometric redshift. All objects with a F814W magnitude above 25.2 are also removed.
    \item The cutout for each source is placed at the estimated source center and extends to approximately five times the half-light radius of the galaxy (the exact formula can be found in \citealt{mandelbaum2012precision}). An additional masking and deconvolution step is applied to the images. The deconvolution takes advantage of the TinyTim point spread function (PSF) estimates \citep{krist201120}, and the masked pixels are replaced with correlated noise.  
\end{itemize}

\section{DRIZZLE Pipeline}\label{app:drizz}

To capture the effects of the \textsc{DrizzlePac} pipeline \textsc{paltas} conducts the following procedure:

\begin{enumerate}
    \item Using \textsc{lenstronomy}, a supersampled raytracing image is generated at twice the resolution of the detector (0.02\arcsec pixel scale). This version of the image does not include the detector noise or the the PSF and is simulated in the sky plane.
    \item Using \textsc{astropy}\footnote{\url{https://www.astropy.org/}}, the supersampled image is mapped to four dithered detector images with half pixel offsets (0.04\arcsec pixel scale), corresponding to the standard four-point dithering strategy \citep{gonzaga2012drizzlepac}. This is done through the World Coordinate System (WCS) of the supersampled image and the dithered images, allowing for geometric distortion coefficients to be included in the mapping. However, in this work we do not include any geometric distortion coefficients. 
    \item These four dithered images each represent the detector output for one exposure. Therefore, we convolve these dithered images with the empirical PSF model and add the expected sky and detector noise.
    \item Using the \textsc{drizzle} package, we drizzle these four images onto the output WCS with the corresponding output 0.03\arcsec pixel scale.
\end{enumerate}

\section{Examples of Subhalo Signal}\label{app:res_an}

\begin{figure*}
    \centering
    \includegraphics[scale=0.205]{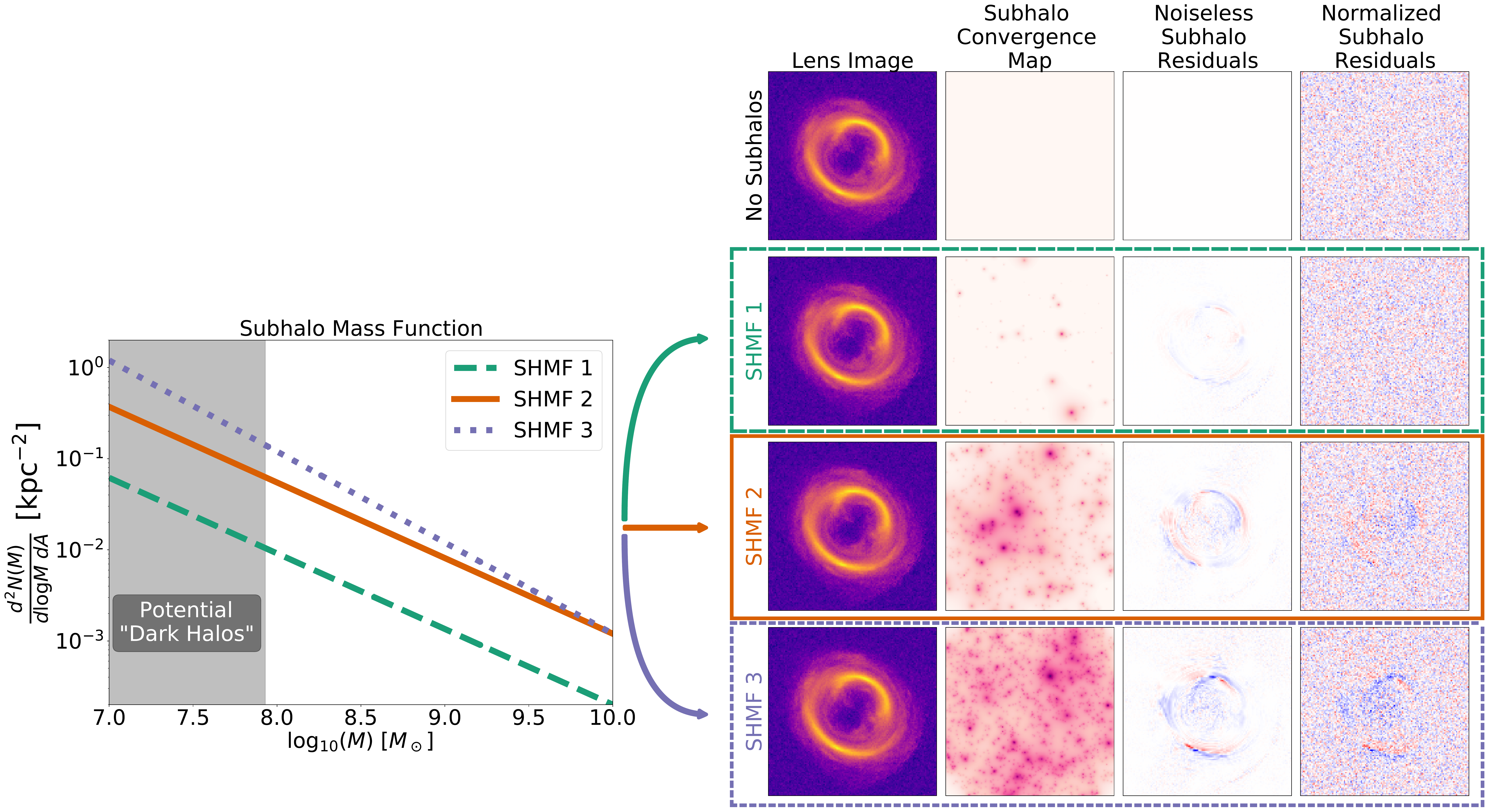}
    \caption{A comparison of the signal imprinted by subhalos for different possible SHMFs. The left hand figure plots the number of subhalos per unit area as a function of mass for the three SHMFs. The right hand figure shows how subhalos drawn from these three functions impact the lens image (left), the convergence induced by subhalos (second from the left),  the noiseless residual signal over a smooth main deflector (second from the right), and the residual signal when normalized by the noise (right). Even a small amount of subhalos is enough to impart some residual signal over the smooth main deflector model, however that signal is subdominant to the noise of the observation. As we increase the slope and normalization of the subhalo mass function, we get more subhalo convergence, and therefore a larger residual signal. However, the detectable signal is localized to the brightest regions of the lens.}
    \label{fig:shmf_toy_varied}
\end{figure*}

\begin{figure*}
    \centering
    \includegraphics[scale=0.205]{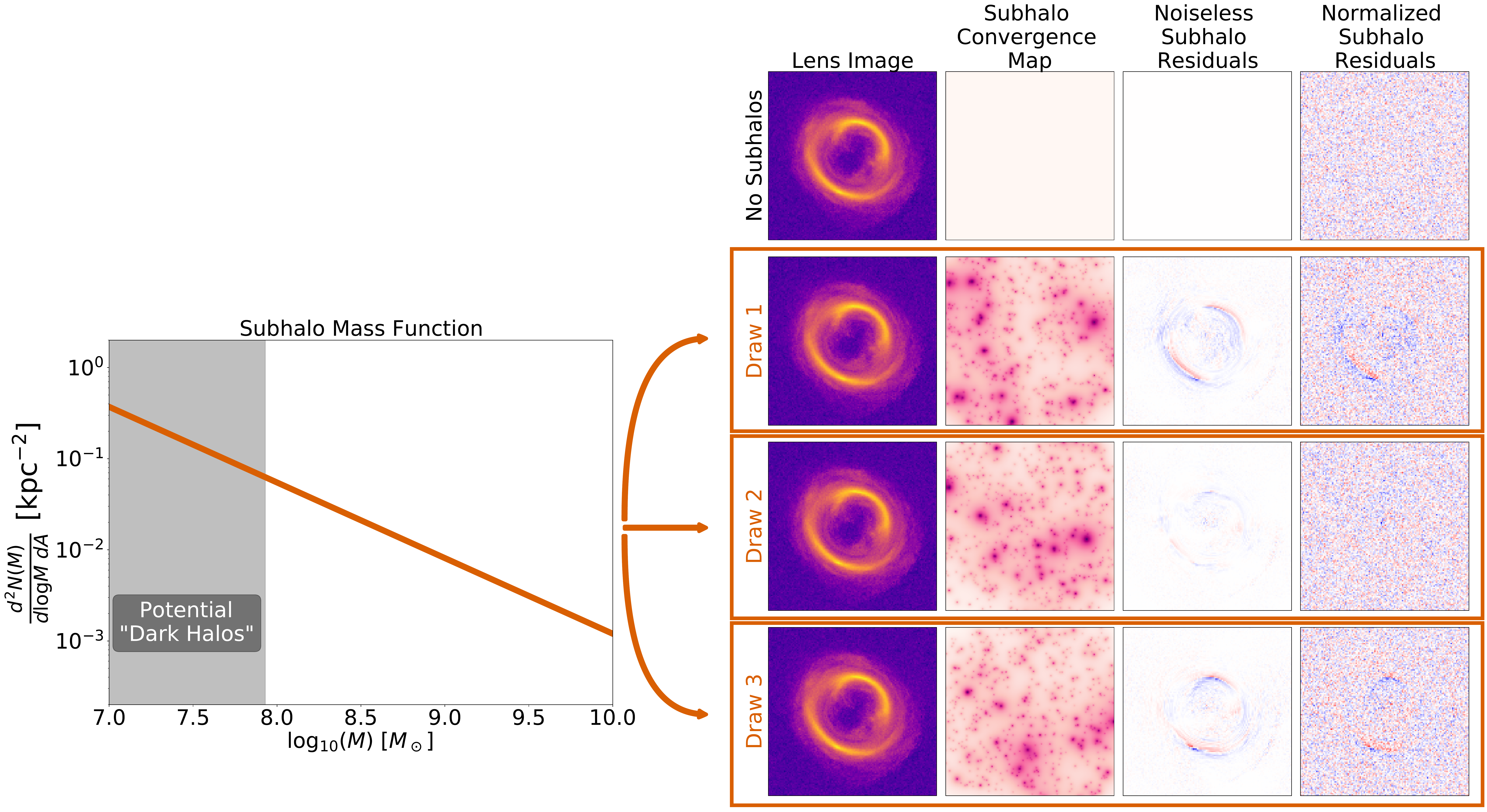}
    \caption{The same plot as in Figure \ref{fig:shmf_toy_varied} but with several draws from the same underlying SHMF. Even for a fixed SHMF, the residual signal can vary drastically depending on how many subhalos overlap with the brightest parts of the image.}
    \label{fig:shmf_toy_fixed}
\end{figure*}

Before training a network to predict the SHMF, it is important to develop intuition for how varying the SHMF impacts the signal in our image. This will help us interpret our network's outputs in Section \ref{sec:perf_train} and the results of our hierarchical analysis in Section \ref{sec:recon_shmf}. To do this, we simulate a set of lens images for different SHMF realizations and model the potential of their main deflector through forward modeling. We then visualize the remaining `residual' lensing signal that cannot be described by the smooth potential of the main deflector. The strength of this residual signal is a direct indicator of the subhalos present in the lensing field. To keep the forward modeling tractable for this example, we make a number of simplifications in this section. We keep the full complexity of the main deflector and subhalos, but we do not render the line-of-sight halos or use the full drizzling and empirical PSF process. Rather, we assume an angular resolution of $0.04 \arcsec$ and a Gaussian PSF with full-width-half-max of $0.04 \arcsec$. We also use a COSMOS source drawn from the same process described in Section \ref{sec:source}, but we assume full knowledge of the source in our simplified model fitting. In the more realistic scenario we tackle in Section \ref{sec:results}, the PSF, the source uncertainty, and the line-of-sight halos will all reduce the strength of the residual subhalo lensing signal. Regardless, for this toy problem, separating out those uncertainties will help focus on the effects of the subhalos.

In Figure \ref{fig:shmf_toy_varied} we show the residual signal generated by three different SHMFs. All three SHMFs follow the parameterization outlined in Section \ref{sec:subhalos}, but with different values for the SHMF normalization and slope. SHMF 1,2, and 3 have a normalization of $\Sigma_\text{sub} = 2.4\times10^{-4},1.2\times10^{-3},$ and $1.2\times10^{-3} \ (\text{kpc})^{-2}$ and a slope of $\gamma_\text{sub} = -1.83,-1.83,$ and $-2.00$ respectively. We show the residual plots for these three SHMFs and for the case of no subhalos. As expected, without subhalos and with perfect knowledge of the source light, the entire lensing signal can be described by our best-fitting main deflector model. As we introduce a small number of subhalos in SHMF 1, we see a small residual signal emerge that cannot be described by our smooth main deflector model ($\chi^2_\nu$: -1.02); however, that signal is nearly subdominant to even the simplified noise model we are using in this toy example. SHMF 2, which has a factor of five larger normalization than SHMF 1, produces a much stronger residual signal that can be visually identified over the noise of the instrument ($\chi^2_\nu$: -1.11). Finally, increasing the steepness of our SHMF, as we do for SHMF 3, introduces many more low-mass subhalos and leads to an even stronger residual signal ($\chi^2_\nu$: -1.20). While the simulations in Figure \ref{fig:shmf_toy_varied} are simplified, they suggest that for large enough normalization and slope, the presence of subhalos in our simulated images should be detectable by our modeling tools. Additionally, we should be able to distinguish between SHMFs with larger slopes and normalizations based on the strength of this residual signal. 

Figure \ref{fig:shmf_toy_varied} shows only one draw per SHMF. If instead we draw multiple times from the same SHMF, as we do for SHMF 2 in Figure \ref{fig:shmf_toy_fixed}, a very different picture emerges. With a fixed amplitude and slope, the convergence maps that are produced looks very similar. However, the exact position of those subhalos with respect to the path of the lensed light leads to very different residual signals. Draw 1 returns a strong residual signal ($\chi^2_\nu$: -1.16), almost comparable with that of SHMF 3 in Figure \ref{fig:shmf_toy_varied}, Draw 2 returns almost no residual signal ($\chi^2_\nu$: -1.05), comparable with SHMF 1 in Figure \ref{fig:shmf_toy_varied}, and Draw 3 is the only one that returns a signal that looks equivalent to what we produced for SHMF 2 in Figure \ref{fig:shmf_toy_varied} ($\chi^2_\nu$: -1.12). The stochastic nature of the residual signal suggests that any modeling technique will return a large uncertainty on our SHMF parameters for an individual lens. This uncertainty is a consequence of attempting to constrain the SHMF, which is a statistical summary of the subhalos, rather than attempting to constrain the thousands of deterministic parameters that describe the subhalos. While the stochastic nature of the residual signal may limit what we can say about an individual lens, it may also allow us to improve our sensitivity to SHMF with very few subhalos. For example, on average draws from SHMF 1 produce residual signals that are subdominant to the noise. But occasionally a draw from SHMF 1 will return a subhalo population that produces a detectable residual signal and is therefore inconsistent with having no subhalos. Hierarchically combining these constraints could therefore allow us to make a meaningful detection of SHMF 1. 

We will explore the signal we can extract for a more realistic population of lenses in Section \ref{sec:recon_shmf}. For now, this toy example leaves us with two takeaways:

\begin{itemize}
    \item The stochastic nature of the residual signal for a fixed SHMF means that we should expect large uncertainties on constraints derived from individual lenses. 
    \item These large uncertainties will require us to conduct hierarchical inference on a population of lenses in order to accurately constrain the SHMF. For more details on this hierarchical inference see Section \ref{sec:hi_inf}.
\end{itemize}

\section{Dataset Parameters}\label{app:params}

\begin{figure}
    \centering
    \includegraphics[scale=0.32]{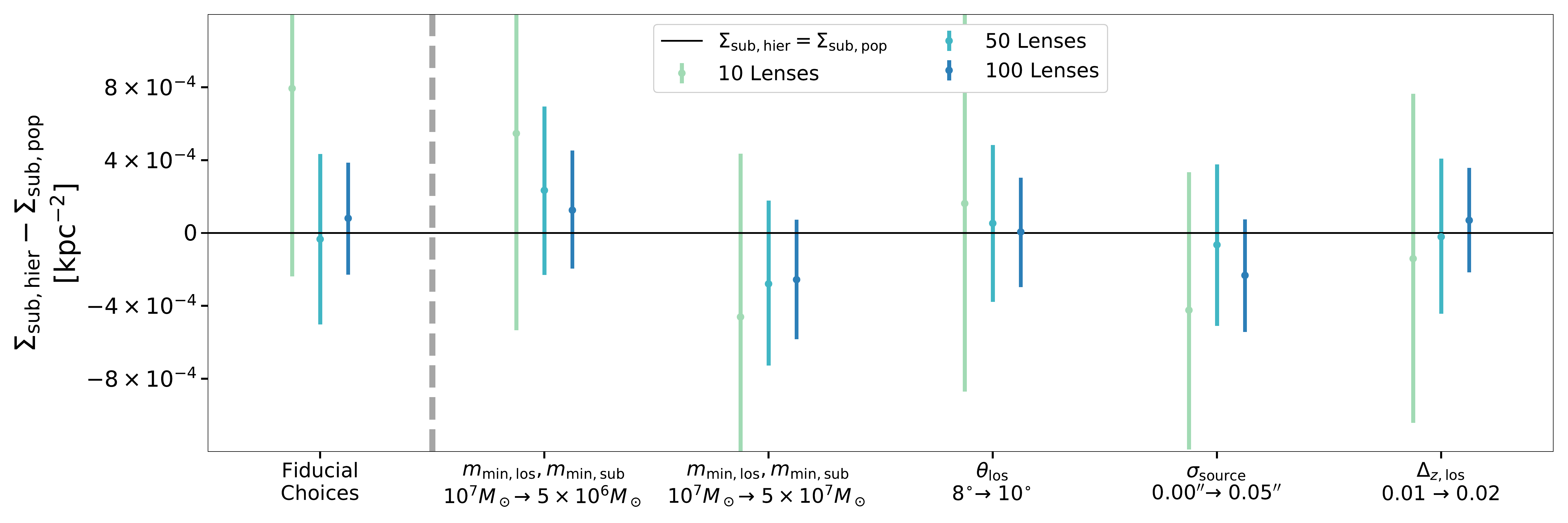}
    \caption{Difference between the inferred population mean, $\Sigma_\text{sub,hier}$, and the true population mean $\Sigma_\text{sub,pop}$ for six test sets. The `Fiducial Choices' test set has the resolution parameters set to the values used throughout this work. The remaining five test sets each change one resolution assumption. The error bars on the inferred $\Sigma_\text{sub,hier}$ show the 68\% interval derived from the hierarchical inference pipeline. All six test sets have a value of $\Sigma_\text{sub,pop} = 2 \times 10^{-3} \ \text{kpc}^{-2}$.}
    \label{fig:res_tests}
\end{figure}

\begin{figure}
    \centering
    \includegraphics[scale=0.32]{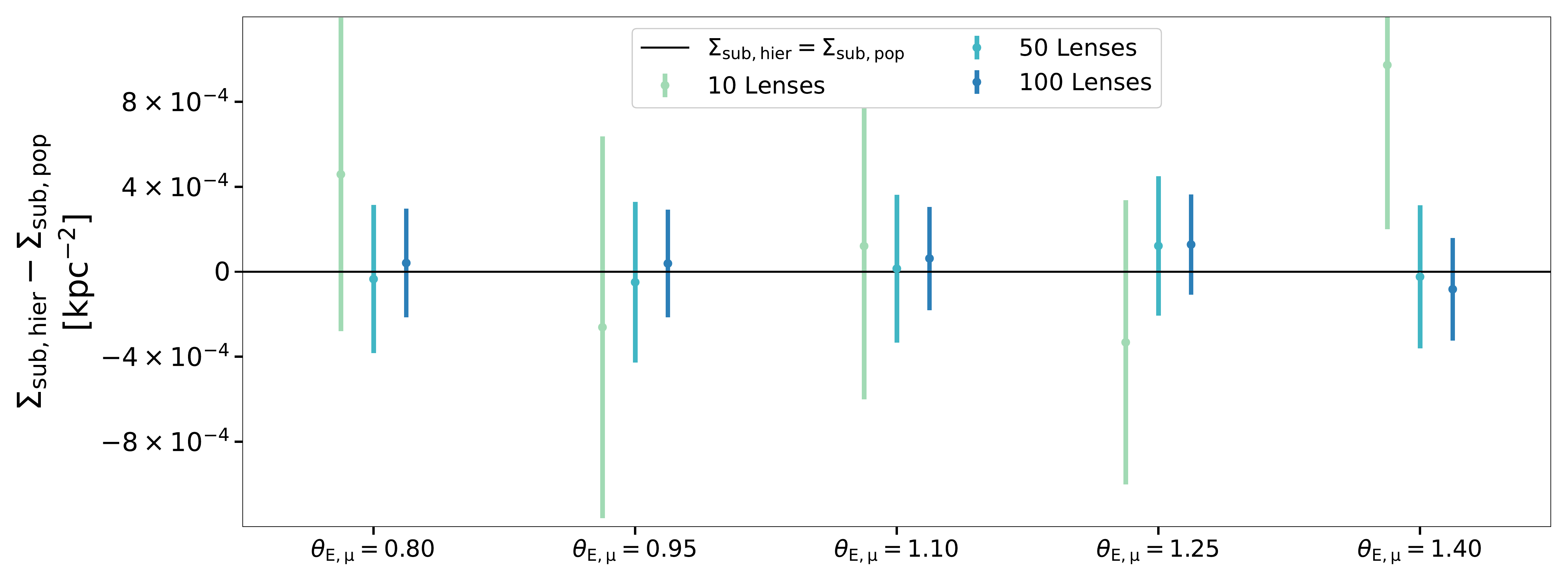}
    \caption{Difference between the inferred population mean, $\Sigma_\text{sub,hier}$, and the true population mean $\Sigma_\text{sub,pop}$ for five test sets with varying, fixed Einstein radius, $\theta_{\text{E},\mu}$. Only the distribution of $\theta_\text{E}$ has been changed, and all other parameters are distributed as in the training set. The error bars on the inferred $\Sigma_\text{sub,hier}$ show the 68\% interval derived from the hierarchical inference pipeline. All four test sets have a value of $\Sigma_\text{sub,pop} = 2.0 \times 10^{-3} \ \text{kpc}^{-2}$.}
    \label{fig:shmf_shift}
\end{figure}

\begin{figure}
    \centering
    \includegraphics[scale=0.32]{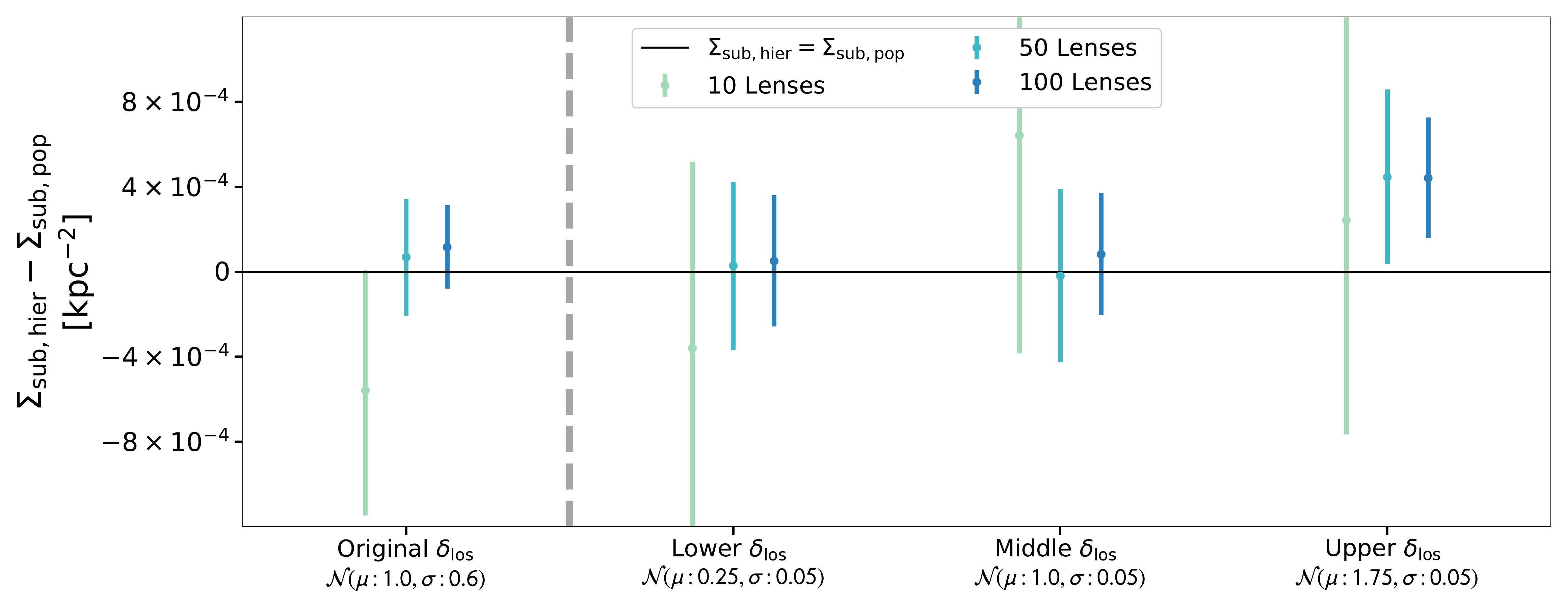}
    \caption{Difference between the inferred population mean, $\Sigma_\text{sub,hier}$, and the true population mean $\Sigma_\text{sub,pop}$ for four test sets with varying distributions of the LOS mass function normalization, $\delta_\text{los}$. The `Original' test set corresponds to test set 7 from the main body. For the three test sets to the right of the grey dashed line, only the distribution of $\delta_\text{los}$ has been changed, and all other parameters are distributed as in test set 7. The error bars on the inferred $\Sigma_\text{sub,hier}$ show the 68\% interval derived from the hierarchical inference pipeline. All four test sets have a value of $\Sigma_\text{sub,pop} = 1.4 \times 10^{-3} \ \text{kpc}^{-2}$.}
    \label{fig:los_shift}
\end{figure}

\begin{figure}
    \centering
    \includegraphics[scale=0.25]{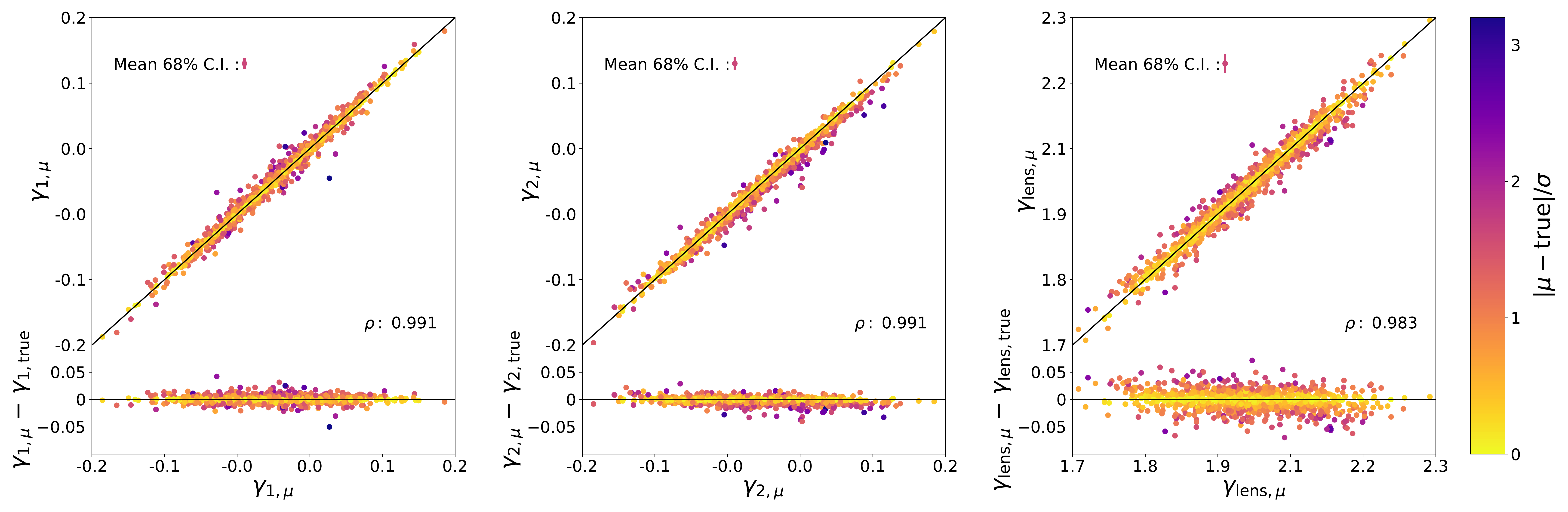}
    \includegraphics[scale=0.25]{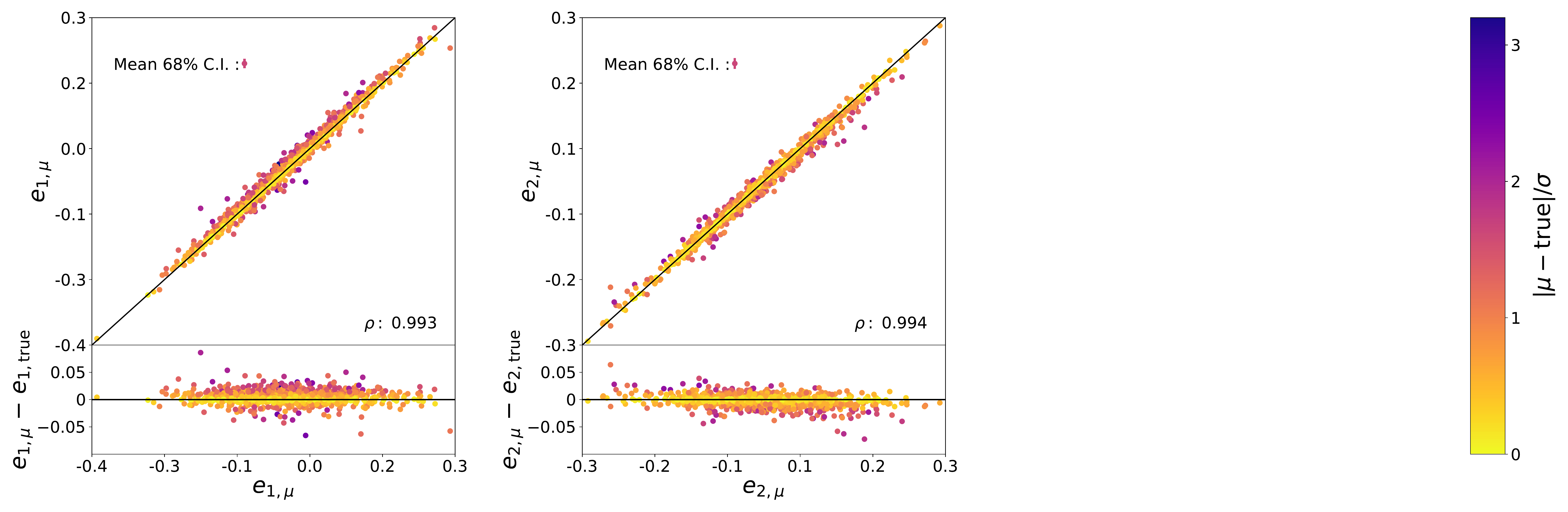}
    \includegraphics[scale=0.25]{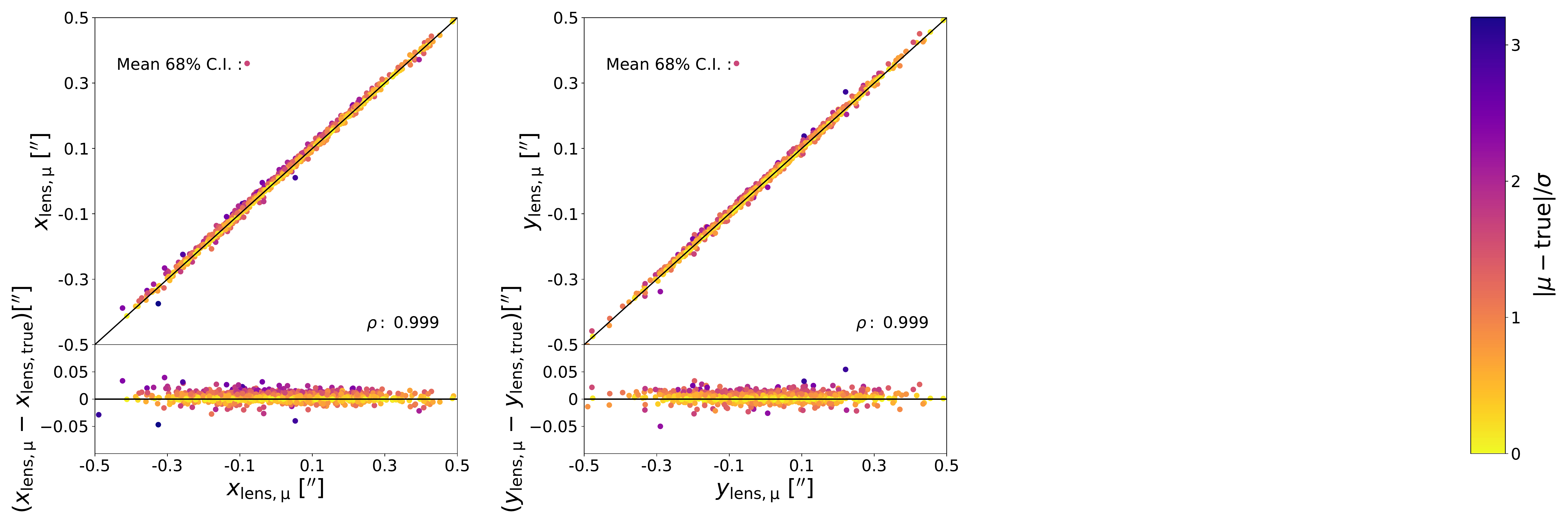}
    \caption{Comparison of the network predicted means and the true input values for all of the lenses in our validation set. The plots show all of the parameters not covered in Figure \ref{fig:confidence_plot}. Going left to right and then top to bottom, these are the x-direction shear, $\gamma_1$, the xy-direction shear, $\gamma_2$, the power-law slope, $\gamma_\text{lens}$, the x-direction ellipticity eccentricity, $e_1$, the xy-direction ellipticity eccentricity, $e_2$, the x-coordinate lens center, $x_\text{lens}$, and the y-coordinate lens center, $y_\text{lens}$. For all of the plots, one point corresponds to the predictions of one lens. The color of the point is defined by the distance between the predicted mean and the truth in units of the predicted standard deviation. The colorbar for all of the plots is the same, and is included on the far right of each row. The upper left region of each plot includes a visualization of the average standard deviation being predicted by the network (equivalent to the 68\% confidence interval). The $\rho$ value printed in the plots represents the Pearson correlation coefficient between the predicted mean and the truth.}
    \label{fig:val_remain}
\end{figure}

For the simulated datasets we use in this work, a number of the parameters associated with the resolution of the simulation are kept fixed. These include the line-of-sight redshift bin widths, the mass boundaries for our subhalos and line-of-sight halos, and the opening angle of our cone. We also assume that the size and intensity of the COSMOS images are large enough to ignore the detector noise in our source model. We test the effect of these choices by generating test datasets with the same parameter distributions as our training dataset but with modified resolution parameters. In particular, we generate one test dataset where we decrease the minimum subhalo and line-of-sight halo mass rendered to $5 \times 10^6 M_\odot$, one test dataset where the minimum mass is increase to $5 \times 10^7 M_\odot$, one test dataset where we increase the opening angle of our cone to 10 degrees, one test dataset where we smooth the source images by a Gaussian psf with a fwhm of $0.05^{\prime\prime}$, and one test dataset where we increase the line-of-sight bin width to 0.02 in redshift space. We run the same hierarchical inference pipeline as in Section \ref{sec:recon_shmf} to see if any of these changes in the simulation choices affect the accuracy of our inference. We only consider 10, 50, and 100 lens analyses since we have found that the 1000 lens analysis can have systematic bias. The results are found in Figure \ref{fig:res_tests} and show that none of the changes in the resolution choices produce an inferred population mean more than one sigma from the true population mean. Each test dataset is statistically independent, so fluctuations within the inference uncertainties are to be expected. This suggests that the results of the analysis in this paper are not significantly impacted by our resolution choices.

Additionally, for this work, we assume a fixed redshift for the source and main deflector. We also assume a fixed main halo mass $m_\text{host}$. The main halo mass is independent of the lensing parameters of the main deflector, and we have absorbed the dependence of the SHMF on host mass into our definition of $\Sigma_\text{sub}$ (see Section \ref{sec:subhalos}). Therefore the main impact of this choice in our simulation is to fix the contribution from our two halo term (see Section \ref{sec:los}). 

\section{Further Hierarchical Inference Tests for the Einstein Radius}\label{app:einst_radius}

As we discuss in Section \ref{sec:sim_datasets}, the test sets have an underlying distribution for the main deflector parameters that is shifted from the training set. The purpose of this shift is to demonstrate that our final inference on the SHMF normalization is not biased when the training set assumptions are violated. This is particularly important for the Einstein radius, since Figure \ref{fig:corner_plot} shows that the network outputs a strong correlation between the inferred Einstein radius and the inferred SHMF normalization. In order to further test our dependence on the assumed distribution of Einstein radii, we have designed 5 additional test sets that have identical distributions to training set except for the Einstein radius which is given a fixed, mean value for all lenses. We vary this fixed value, $\theta_{\text{E},\mu}$, to span the two sigma contours of the training set prior. The inferred SHMF normalization can be seen in Figure \ref{fig:shmf_shift}. We find no evidence of systematic bias in the inferred mean SHMF normalization as we vary the Einstein radius of the lens population. If our network were only sensitive to the Einstein radius and not the underlying signal of low-mass halos, we would not be able to reconstruct an accurate and precise mean for the SHMF normalization on these test sets. Therefore, we conclude that the information our network is extracting from the images goes beyond the observed radius of the Einstein ring.

\section{Line-of-Sight Halos Normalization}\label{app:delta_los}

Throughout this work, we have made a fixed assumption about the distribution of line-of-sight halo normalizations, $\delta_\text{los}$. As we discuss in more detail in Section \ref{sec:los}, previous studies have suggested that the line-of-sight halos can produce deflections that are comparable to the subhalos. To better understand how constraining our assumed line-of-sight halo distribution is to our inference, we generate three test sets with much narrower line-of-sight distributions that span the `lower', `middle', and `upper' regions of the training distribution. This corresponds to a distribution of $\delta_\text{los} \sim \mathcal{N} (\mu : 0.25, \sigma:0.05), \ \mathcal{N} (\mu : 1.0, \sigma:0.05), \  \mathcal{N} (\mu : 1.75, \sigma:0.05)$ respectively. The goal of these test sets is to probe how a systematic error in the average line of sight translates to a bias in the inferred subhalo mass function normalization, $\Sigma_\text{sub,hier}$. To help make a direct comparison, we fix all of the other parameters for these test sets to the same values used for training set 7 (see Section \ref{sec:sim_datasets} for more details). As with the tests in Appendix \ref{app:params}, we only consider 10, 50, and 100 lens analyses. The results of the hierarchical inference are shown in Figure \ref{fig:los_shift}. For the lower and middle test sets, the change in the line-of-sight distribution does not shift the inferred mean more than one sigma from the true population mean. Only for the `upper' test set is the shift more than one sigma, and the final result is still within the 95\% confidence interval. We conclude that, for the shifts in the $\delta_\text{los}$ distributions we explore here, the induced bias in $\Sigma_\text{sub,hier}$ is sub-dominant to the uncertainties for 10, 50, and 100 lenses.  

\section{Validation Predictions for Remaining Parameters}\label{app:val_remain}

In Figure \ref{fig:val_remain}, we present the predicted mean values as a function of the true values for the lens parameters not shown in Figure \ref{fig:confidence_plot}. The points shown here span the entire validation set. As with the the Einstein radius, the network's predictions for the remaining main deflector parameters are strongly correlated with the truth $(\rho >= 0.98)$.

\end{document}